\documentclass[preprint,aip,jcp]{revtex4-1}
\usepackage{epsfig}
\usepackage{textcomp}
\usepackage{graphicx}
\usepackage{amsmath}
\usepackage{url}
\usepackage[normalem]{ulem}

\usepackage{color}
\usepackage{xr}
\newcommand*{\citen}[1]{%
  \begingroup
    \romannumeral-`\x 
    \setcitestyle{numbers}%
    \cite{#1}%
  \endgroup
}

\definecolor{Myblue}{rgb}{0.3,0.3,1.0}

\usepackage{rotating}
\begin{document}
\title{Near dissociation states for H$_2^+$--He on MRCI and FCI
  potential energy surfaces}

\author{Debasish Koner}
\affiliation{Department of Chemistry, University of Basel,
Klingelbergstrasse 80, CH-4056 Basel, Switzerland}

\author{Juan Carlos San Vicente Veliz}
\affiliation{Department of Chemistry, University of Basel,
Klingelbergstrasse 80, CH-4056 Basel, Switzerland}

\author{Ad van der Avoird} \affiliation{Theoretical Chemistry,
  Institute for Molecules and Materials, Radboud University
  Nijmegen,Nijmegen, The Netherlands}

\author{Markus Meuwly}
\email[]{m.meuwly@unibas.ch}
\affiliation{Department of Chemistry, University of Basel,
Klingelbergstrasse 80, CH-4056 Basel, Switzerland}

\date{\today}

\begin{abstract}
A new analytical potential energy surface (PES) has been constructed
for H$_2^+$--He using a reproducing kernel Hilbert space (RKHS)
representation from an extensive number of {\it ab initio} energies
computed at the multi-reference and full configuration interaction
level of theory. For the MRCI PES the long-range interaction region of
the PES is described by analytical functions and is connected smoothly
to the short-range interaction region, represented as a RKHS. All
ro-vibrational states for the ground vibrational and electronic state
of H$_2^+$--He are calculated using two different methods to determine
quantum bound states. Comparing transition frequencies for the
near-dissociation states for {\it ortho-} and {\it para} H$_2^+$--He
allows assignment of the 15.2 GHz line to a $J=2$ $e/f$ parity doublet
of {\it ortho-}H$_2^+$--He whereas the experimentally determined 21.8
GHz line is only consistent with a $(J=0)$ $\rightarrow$ $(J=1)$ $e/e$
transition in {\it para-}H$_2^+$--He.
\end{abstract}

\maketitle

\section{Introduction}
The interaction between ions and neutral atoms or molecules is of
central importance in atmospheric and astronomical processes and
environments. Prominent species in the interstellar environment
include H$_3^+$, CH$_2^+$ , HCO$^+$ and N$_2$H$^+$, among
others.\cite{tielens:2013} Additionally, ions are also considered to
play an important role in the formation of atmospheric
aerosols.\cite{hao:2015}\\

\noindent
Very recently,\cite{stutzki:2019} the HeH$^+$ ion, which was the first
molecule of the primordial universe,\cite{zyg98:151} has been detected
in interstellar space and means for the direct detection of H$_2^+$
have been discussed.\cite{black:2012} However, although H$_2^+$ is
most likely formed and present in space, e.g. through the HeH$^+$ + H
$\rightarrow$ H$_2^+$ + He reaction,\cite{stutzki:2019} (which is
believed to be the first atom-diatom reaction in the
universe\cite{lep02:R57}) collisions with H and H$_2$ are also
important loss channels of the ion. Nevertheless, with H$_2^+$ present
in the interstellar medium, it is also likely that the H$_2^+$--He
complex is formed. Hence, H$_2^+$--He plays an important role already
in the early stages of the Molecular Universe.\\

\noindent
The interaction between He and H$_2^+$ is also important for the
rotational cooling of H$_2^+$ through collisions with Helium as the
buffer gas.\cite{schiller:2017} This is an attractive way to generate
translationally and internally cold H$_2^+$ ions suitable for
precision measurements.\cite{schiller:2007,schiller:2014} With even
further increased precision and quantum state control of the ions,
fundamental natural constants such as the ratio of the electron to the
proton mass, $m_{\rm e} / m_{\rm p}$, can be determined with
unprecedented accuracy.\\

\noindent
Another process of interest which has been recently investigated is
the Penning ionization of $^3$S excited He colliding with
H$_2$.\cite{klein:2017} This process produces H$_2^+$--He with
sufficient energy to dissociate into ground state and rovibrationally
excited He and H$_2^+$ fragments. Such rovibrationally inelastic
half-collisions are particularly sensitive to the long-range part of
the intermolecular potential, which is dominated by polarization
interactions induced by the charge and quadrupole of
H$_2^+$. Furthermore, several long-range states for H$_2^+$--He have
been characterized from microwave spectroscopy and by using electric
field extraction.\cite{car96:395,gam02:6072} However, the
interpretation of these spectra has remained elusive, in part due to
the limited accuracy of the available potential energy surfaces.\\

\noindent
In the past, several PESs have been constructed at different levels of
theory to investigate the spectroscopy and dynamics of the H$_2^+$--He
complex.\cite{jos87:704,fal99:117,meu99:3418,pal00:1835,ram09:26,faz12:244306}
To characterize spectral transitions in the microwave region, an
accurate long-range potential is required.\cite{fal99:117,meu99:3418}
However, the level of theory used for the electronic structure
calculations in these earlier efforts was rather modest by today's
standards. Full configuration interaction (FCI) with the cc-pVQZ basis
has been used more recently but no explicit analytical representation
was included.\cite{ram09:26} Later, using the {\it ab initio} data of
Ref. \citen{ram09:26} a new PES was constructed by including an
explicit analytical formula only for the diatomic
potentials.\cite{faz12:244306}\\

\noindent
In the present work high-level electronic structure methods combined
with advanced representation techniques for global potential energy
surfaces and accurate representation of the long-range potential are
used. With these PESs quantum calculations of all bound states of
H$_2^+$--He with H$_2^+$ in its ground electronic and vibrational
state are then carried out. First, the computational methods are
presented, followed by the discussion of the bound states computed and
their interpretation in view of the near-dissociative states.\\

\section{Computational Methods}

\subsection{The Potential Energy Surfaces}
Two different levels of theory - a) multi reference configuration
interaction level including the Davidson correction
(MRCI+Q)\cite{wer88:5803,kno88:514} with the augmented Dunning-type
correlation consistent polarize hexaple zeta
(aug-cc-pV6Z)\cite{wil96:339} basis set and b) full configuration
interaction \cite{kno84:315,kno89:75} with the augmented Dunning
type correlation consistent polarized quintuple zeta
(aug-cc-pV5Z)\cite{dun89:1007,woo94:2975} basis set - are used in the
present work to calculate the {\it ab initio} energies. Initial
orbitals for the MRCI calculations were obtained using the complete
active space self-consistent field
(CASSCF)\cite{wen85:5053,kno85:259,wer80:2342} method with three $1s$
orbitals of H and He in the active space. The Molpro\cite{molpro}
software was used to perform all electronic structure calculations.\\

\noindent
The grids for the {\it ab initio} energy calculations are set up in
Jacobi coordinates ($R, r, \theta$).  Here, $r$ is the H$_2^+$ bond
length, $R$ is the distance between He and the center of mass of the
H$_2^+$ ion, and $\theta$ is the angle between $\vec{r}$ and
$\vec{R}$. The angular grid is defined by Gauss-Legendre quadrature
points chosen in the range between $0 \leq \theta \leq 90^\circ$ given
the spatial symmetry of the system. Details of the angular and radial
grids for the MRCI+Q and FCI calculations are given in Tables
S1 and S2 in the supporting
information.\\

\noindent
The complete adiabatic surface for H$_2^+$--He can be expressed as a
many-body expansion\cite{var88:255}
\begin{equation}
\begin{split}
V_{\rm HeHH'}({r_{\rm HeH}},{r_{\rm HeH'}},{r_{\rm HH'}}) & =
V^{(1)}_{\rm He} + V^{(1)}_{\rm H} + V^{(1)}_{\rm H^+} \\ & +
V^{(2)}_{\rm HeH^+}(r_{\rm HeH}) + V^{(2)}_{\rm HeH'^+}(r_{\rm HeH'})
+ V^{(2)}_{\rm HH'^+}(r_{\rm HH'}) \\ & + V^{(3)}({r_{\rm
    HeH}},{r_{\rm HeH'}},{r_{\rm HH'}}),
\end{split}
\label{mb}
\end{equation}
where ${r_{\rm HeH}},{r_{\rm HeH'}}$ and ${r_{\rm HH'}}$ are the
distances between the respective atoms, and $V_{\rm HeHH'}({r_{\rm
    HeH}},{r_{\rm HeH'}},{r_{\rm HH'}})$ is the total energy of the
triatomic system at the corresponding geometry. The $V_i^{(1)}$ are
the atomic energies , whereas the $V_i^{(2)}(r_i)$ and
$V^{(3)}({r_{\rm HeH}},{r_{\rm HeH'}},{r_{\rm HH'}})$ are the two- and
three-body interaction energies, respectively, at corresponding
configurations.\\

\noindent
In general, two body interaction energies, i.e., the diatomic
potential, for a molecule AB can be expressed
as\cite{agu92:1265,faz12:244306}
\begin{equation}
 V^{(2)}_{\rm AB}(R_{{\rm AB}}) = \frac{c_0e^{-\alpha_{{\rm AB}}
     R_{{\rm AB}}}}{R_{{\rm AB}}} + \sum_{i=1}^{M}c_i\rho^i_{{\rm AB}}
 + V_{\rm long}(\tilde r),
 \label{ap2}
\end{equation}
with $c_0 > 0$ to ensure $V_{\rm AB}(R_{{\rm AB}}) \rightarrow \infty$
at $R_{{\rm AB}} \rightarrow 0$ and $\rho_{{\rm AB}} = R_{{\rm
    AB}}e^{-\beta_{{\rm AB}}^{(2)}R_{{\rm AB}}}$. The long range part,
$V_{\rm long}(\tilde r)$, can be written as\cite{fal99:117}
\begin{equation}
V_{\rm long}(\tilde r) = - \frac{\alpha_d q^2}{2\tilde r^4} -
\frac{\alpha_q q^2}{2\tilde r^6} - \frac{\alpha_o q^2}{2\tilde r^8} -
\frac{\beta_{ddq} q^3}{6\tilde r^7} - \frac{\gamma_d q^4}{24\tilde
  r^8},
 \label{dal}
\end{equation}
where $q$ is the charge, and $\alpha_d, \alpha_q$ and $\alpha_o$ are
the dipole, quadrupole and octopole polarizabilities for H and He,
respectively. $\beta_{ddq}$ and $\gamma_d$ are the first and second
hyperpolarizabilities, respectively. The values for the
polarizabilities of He and H are taken from
Refs. \citen{fal99:117,bis95:15} and $\tilde r$ is defined
as\cite{vel08:084307}
\begin{equation}
 \tilde r = r + r_l \exp^{(-(r-r_e))},
\end{equation}
to remove the divergence of the long range terms at short H-H and H-He
separations. Here, $r_l$ is a distance parameter and $r_e$ is the
equilibrium bond distance of the diatomic molecule. The parameters
used in this work to obtain the diatomic potentials are given in Table
\ref{tab:params}.\\

\begin{table}[ht]
\caption{Parameters used in the diatomic potentials. All values are in
  atomic units.}
\begin{tabular}{l|r|r}
    \hline
    \hline
         & H & He  \\
    \hline
    \hline
    Dipole polarizability $\alpha_d$&4.5&1.384\\
    Quadrupole polarizability $\alpha_q$ &15.0&2.275 \\
    Octopole polarizability $\alpha_o$&131.25&10.620 \\
    First hyperpolarizability $\beta_{ddq}$&159.75&20.41 \\
    Second hyperpolarizability $\gamma_d$&1333.125&37.56 \\
    \hline
    & H$_2^+$ & HeH$^+$  \\
    \hline
    \hline
    $r_l$ &10.0&8.0\\
    $r_{\rm eq}$&2.005815&1.4633\\
    \hline
    \hline
  \end{tabular}
  \label{tab:params}
\end{table}

\noindent
The linear parameters $c_i$ and the nonlinear parameters $\alpha_{{\rm
    AB}}$ and $\beta_{{\rm AB}}^{(2)}$ in Eq. \ref{dal} are determined
by fitting the expression with the {\it ab initio} energies using the
Levenberg-Marquardt nonlinear multidimensional fitting
method.\cite{pre92} The optimized linear and nonlinear parameters for
the diatomic potentials calculated via fitting are given in Tables
S3 and S4.\\

\noindent
The three-body interaction energies, $V^{(3)}({r_{\rm HeH}},{r_{\rm
    HeH'}},{r_{\rm HH'}}) \equiv V^{(3)}(r, R, \theta)$ are calculated
from Eq. \ref{mb}. For a particular configuration of He-H$_2^+$,
$V^{(3)}(r, R, \theta)$ can be calculated using the reproducing kernel
Hilbert space\cite{ho96:2584} (RKHS) approach.\\

\noindent
The procedure for computing the analytical energy of a given
configuration from a set of known {\it ab initio} energies is briefly
described here.  According to the RKHS theorem, the value of a
function $f({\bf x})$ can be evaluated from a set of known values
$f({\bf x}_i)$ at positions ${\bf x}_i$ as a linear combinations of
kernel products
\begin{equation}
\tilde{f}({\bf{x}}) = \sum_{i=1}^{N} c_i K ({\bf{x}}, {\bf{x}}_i),
\end{equation}
where $c_i$ are the coefficients and $K({\bf{x}}, {\bf{x}}_i)$ are the
reproducing kernels. The coefficients are calculated from the known
values by solving a set of linear equations
\begin{equation}
 f({\bf{x}}_j)= \sum_{i=1}^{N} c_i K ({\bf{x}}_i, {\bf{x}}_j).
\end{equation}
Here it is worth mentioning that the RKHS approach exactly reproduces
the input data at the reference points. The derivatives of
$\tilde{f}({\bf{x}})$ can be calculated analytically from the kernel
functions $K({\bf{x}}, {\bf{x}}')$. For a multidimensional function
the $D$-dimensional kernel can be constructed as the product of $D$
1-dimensional kernels $k(x, x')$
\begin{equation}
 K({\bf{x}}, {\bf{x}}') = \prod_{d=1}^{D}k^{(d)}(x^{(d)}, x'^{(d)}),
\end{equation}
where $k^{(d)}(x^{(d)}, x'^{(d)})$ are the 1-dimensional kernels for
$d$-th dimensions.\\

\noindent
For the radial dimensions ($r$ and $R$) a reciprocal power decay
kernel
\begin{equation}
\label{k24}
 k^{[2,4]}(x,x') = \frac{2}{15}\frac{1}{x^5_{>}} -
 \frac{2}{21}\frac{x_<}{x^6_>},
\end{equation}
is used in the present work where, $x_>$ and $x_<$ are the larger and
smaller values of $x$ and $x'$. The value of this kernel smoothly
decays to zero according to $x^{-4}$ as the leading term in the
asymptotic region, which gives the correct long-range behavior for
atom-diatom type interactions. For the angular dimension, a Taylor
spline kernel
\begin{equation}
 k^{[2]}(z,z') = 1 + z_<z_> + 2z^2_<z_> - \frac{2}{3}x^3_<,
\end{equation}
is used, where $z_>$ and $z_<$ are analogous to $x_>$ and $x_<$. Here,
the variable $z$ is defined as
\begin{equation}
z = \frac{1 - {\rm cos} \theta}{2},
\end{equation}
so that the values of $z$ are always in the interval [0,1].\\

\noindent
Finally, the 3-dimensional kernel is
\begin{equation}
\label{3dk}
 K({\bf{x}}, {\bf{x}}') = k^{[2,4]}(R,R')k^{[2,4]}(r,r')k^{[2]}(z,z'),
\end{equation}
where, ${\bf{x}}, {\bf{x}}'$ are $(R, r, z)$ and $(R', r', z')$,
respectively. A computationally efficient toolkit is used in this work
to calculate the coefficients and in evaluating the
function\cite{unk17:1923}. Adding a small regularization parameter
(here $\lambda = 10^{-20}$ for the MRCI+Q data) to the diagonal
elements provides additional numerical stability. In practice,
$\lambda$ is increased until a regular solution is obtained for the
inversion. For FCI no regularization is required.\\

\noindent
To represent the long range part of the H$_2^+$--He interaction
the analytical form from Ref. \citen{fal99:117}
 \begin{equation}
 \begin{split}
  V_{\rm long}(R, r, \theta) = & - \frac{\alpha_d q^2}{2R^4} -
  \frac{\alpha_q q^2}{2R^6} - \frac{\alpha_0 q^2}{2R^8} -
  \frac{\beta_{ddq} q^3}{6R^7} - \frac{\gamma_d q^4}{24R^8}
  \\ &-\frac{3\alpha_dq\Theta(r)P_2(\rm{cos}\theta)}{R^6}
  -\frac{5\alpha_dq\Phi(r)P_4(\rm{cos}\theta)}{R^8}
  -\frac{6\alpha_qq\Theta(r)P_2(\rm{cos}\theta)}{R^8}
  \\ &-\frac{C_6^0(r) + C_6^2(r)P_2(\rm{cos}\theta)}{R^6} -
  \frac{C_8^0(r) + C_8^2(r)P_2(\rm{cos}\theta)}{R^8}
  \end{split}
  \label{tl}
 \end{equation}
is used. Here, the first five terms represent the charge+induced
multipole interactions, the sixth term represents the
dipole+quadrupole induction interaction and the seventh and eighth
terms represent the higher order induced-dipole+hexadecapole and
induced-quadrupole+quadrupole interactions, respectively.  Here,
$\Theta(r)$ and $\Phi(r)$ are the quadrupole and hexadecapole moments
of H$_2^+$, respectively. The last two terms in Eq. \ref{tl} are the
contributions from dispersion interactions. The $r-$dependence of the
moments and dispersion coefficients is included by representing them
as a second degree Taylor series. All parameters used in Eq. \ref{tl}
are those from Ref. \citen{fal99:117}. The parameters for He are
given in Table \ref{tab:params}.\\

\begin{figure}
\includegraphics[scale=1.4]{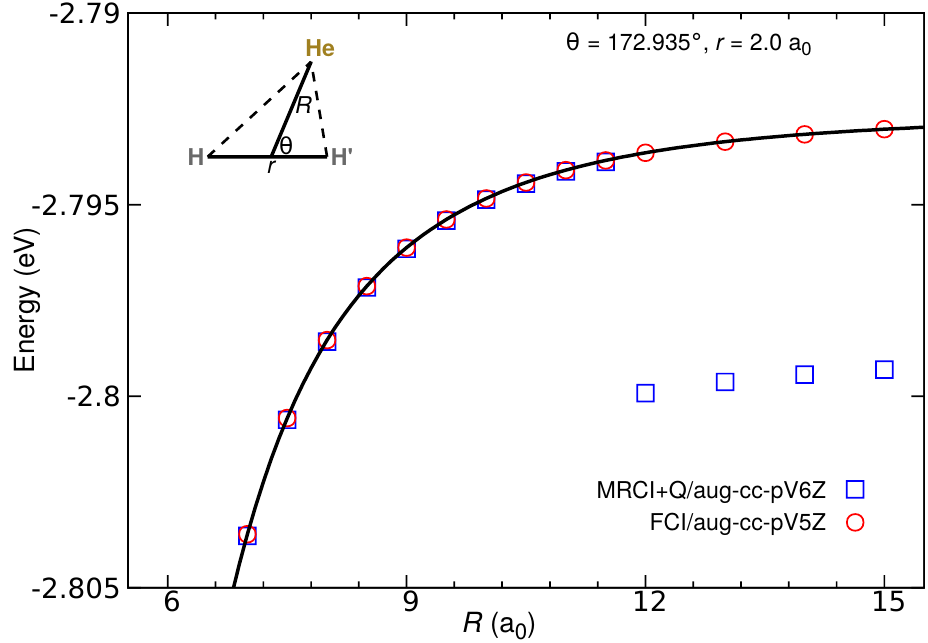}
\caption{MRCI+Q/aug-cc-pV6Z and FCI/aug-cc-pV5Z energies as a function
  of $R$ for fixed $\theta = 172.935^\circ$ and $r = 2.0$ a$_0$. The
  black line is the RKHS interpolation of the FCI energies}
\label{fig1}
\end{figure}

\begin{figure}
\includegraphics[scale=1.6]{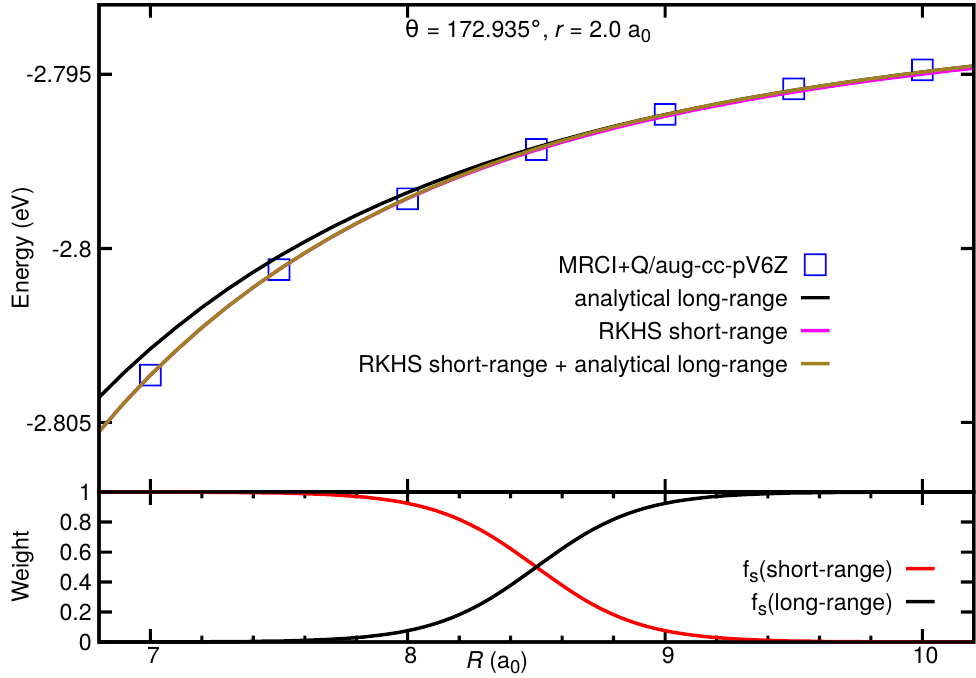}
\caption{Upper panel: Energies obtained from MRCI+Q/aug-cc-pV6Z, RKHS,
  analytical long-range and RKHS+analytical are plotted as a function
  of $R$ for fixed $\theta = 172.935^\circ$ and $r = 2.0$ a$_0$. Lower
  panel: the weights for the short range interaction energies and long
  range interaction energies as a function of $R$.}
\label{fig1a}
\end{figure}

\noindent
For configurations with $R \gtrsim 9.0$ a$_0$ the MRCI+Q/aug-cc-pV6Z
calculations are discontinuous along the $R-$coordinate, see Figure
\ref{fig1}, where MRCI+Q/aug-cc-pV6Z energies are given fixed values
of $\theta$ and $r$.  For large values of $R$, MRCI+Q/aug-cc-pV6Z
energies are discontinuous which originates from the Davidson
correction of the MRCI energies because the order of the states along
a potential energy scan can swap. This then leads to discontinuities
in the Davidson-corrected energies. Hence, for the long range part of
the PES the explicit analytical long-range expression (see
Eq. \ref{tl}) was used to construct the full 3D PES which is referred
to as MRCI+Q+LR in the following. In order to smoothly connect the
short- and long-range parts of the MRCI+Q PES a Fermi (switching)
function is used, see Figure \ref{fig1a}:
\begin{equation}
  f_s(R) = \frac{1}{{\rm exp}\frac{R-R_0}{\delta R}+1}
\label{eq:fermi}
\end{equation}
where $R_0 = 8.5$ a$_0$ and $\delta R = 0.2$ a$_0$.  The function has
a value of 0.5 at $R = 8.5$ a$_0$. The total potential $(V_{\rm tot})$
is then calculated as
\begin{equation}
 V_{\rm tot} = f_s V_{\rm RKHS} + (1-f_s)V_{\rm long},
\label{sfunc}
\end{equation}
where $V_{\rm RKHS}$ is the short range part of the interaction
potential obtained from RKHS interpolation using the many body
expansion.\\

\noindent
Full CI calculations are smooth out to $R \sim 50$ a$_0$, contrary to
MRCI+Q, see Figure \ref{fig1}. Hence, the full 3-dimensional PES was
also calculated using FCI using a somewhat smaller basis set, i.e.,
aug-cc-pV5Z. This PES, called FCI in the following, was again
represented as a RKHS. Although the FCI energies are smooth in the
$R-$long range, a third PES (FCI+LR) was constructed by using the same
long range expression used for the MRCI+Q+LR PES. For the FCI+LR PES
the parameter values in the switching function were $R_0 = 13.5$ a$_0$
and $\delta R = 0.25$ a$_0$ in Eq. \ref{eq:fermi}.\\

\subsection{Bound state calculations}
Ro-vibrational bound state calculations for different $J$ states with
$e$ and $f$ symmetries are carried out in scattering coordinates using
the 3D discrete variable representation (DVR) method with the DVR3D
program suite.\cite{ten04:85} The radial Gauss-Laguerre quadrature
grids consist of 86 and 32 points along $R$ and $r$ coordinates,
respectively. For the Jacobi angle $\theta$, a grid of 36
Gauss-Legendre points was used and for the radial grids ($r$, $R$) the
wavefunctions were constructed using Morse oscillator functions. For
the diatom (H$_2^+$), $r_e = 2.5$ a$_0$, $D_e = 0.1026$ $E_{\rm h}$
and $\omega_e = 0.018$ $E_{\rm h}$ are used and with these parameters
the $r-$grid covered points between 0.92 to 3.8 a$_0$. As the wave
functions for the near-dissociation states need to cover large values
along $R$, the corresponding values were $R_e = 11.5$ a$_0$, $D_e =
0.08$ $E_{\rm h}$, and $\omega_e = 0.00065$ $E_{\rm h}$ which defined
the $R$ grid between 1.82 and 20.87 a$_0$. The $r_2$
embedding\cite{ten04:85} is used to calculate the rotationally excited
states, where the $z-$axis is parallel to $R$ in body-fixed Jacobi
coordinates. For the $J > 0$ calculations, the Coriolis couplings are
included. In the $r_2$ embedding, calculations with $ipar = 1$ and 0
correspond to the $ortho$ and $para$ H$_2^+$, respectively. The $e$
and $f$ symmetries are assigned by the parity operator $p$.\\

\noindent
Another method by which we calculated the bound states is the
coupled-channels variational method (CCVM). It is similar to a
coupled-channels (CC) scattering calculation, but instead of
propagating the radial coordinate $R$ to solve the CC differential
equations it uses a basis also in $R$ and obtains the desired number
of eigenstates of the Hamiltonian matrix with the iterative Davidson
algorithm \cite{davidson:75}. For the angular motion of H$_2^+$ in the
H$_2^+$--He complex we used a free rotor basis with $j_{H_2^+}$
ranging from 0 to 14 (or 16, in tests). The basis in the H$_2^+$
vibrational coordinate $r$ contains the $v = 0 - 7$ eigenfunctions of
the free H$_2^+$ Hamiltonian for $j_{H_2^+} = 0$ on a grid of 110
equidistant points with $r = 0.25 - 5.5~a_0$. The basis in $R$ was
obtained by solving a one-dimensional (1D) eigenvalue problem with the
radial kinetic energy and a potential $V_{\rm eff}(R)$. This potential
is a cut through the full 3D potential of He-H$_2^+$ with $\theta$ and
$r$ fixed at the equilibrium values, to which we added a term linear
in $R$ with a slope that was variationally optimized by using the $R$
basis in full 3D calculations of the lower He-H$_2^+$ levels. The 1D
radial eigenvalue problem was solved with sinc-DVR
\cite{groenenboom:93b} on a 357-point grid with $R = 2 - 50~a_0$. In
order to converge also near-dissociative states we finally included
120 radial basis functions in the 3D full direct product basis.\\

\section{Results and Discussion}

\subsection{Quality of the PESs}
\begin{figure}
\includegraphics[scale=0.74]{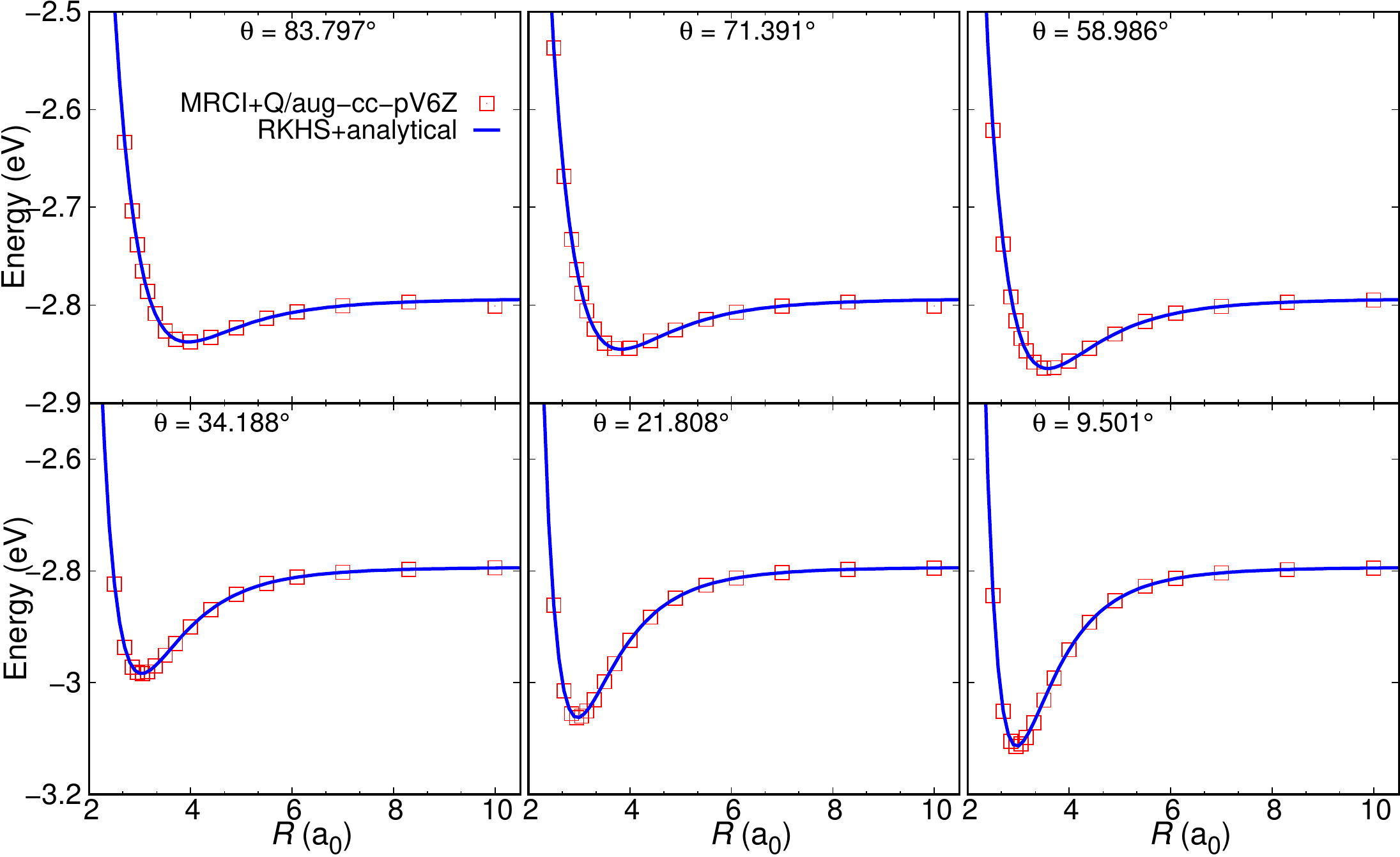}
\caption{The analytical energies obtained from MRCI+Q+LR PES (solid
  lines) and the MRCI+Q/aug-cc-pV6Z {\it ab initio} energies as a
  function of $R$ for several Jacobi angles and at fixed $r = 2.0$
  a$_0$.}
\label{fig2}
\end{figure}

First, the quality of the {\it ab initio} calculations and their RKHS
representation is considered. In Figure \ref{fig2} the analytical
energies are compared with the {\it ab initio} energies for a few
selected Jacobi angles at $r = 2.0$ a$_0$ for the MRCI+Q+LR PES.  A
similar comparison is also shown for the FCI PES in Figure
S1. Excellent agreement between the two sets of data is
found, see Figure \ref{fig2}. Figure S2 presents the
contour plot of the analytical energies for the He-H$_2^+$ system for
$r = 2.0$ a$_0$.\\

\noindent
The quality of the RKHS representation of the MRCI+Q+LR and FCI PES is
reported in Figure S3. For the grid points used to
generate the RKHS representation, the agreement between reference
points and the reproducing kernel is excellent with $R^2$ values of
$(1-2\times10^{-9})$ and $(1-3\times10^{-12})$ for MRCI+Q+LR and FCI
PESs, respectively. In addition, {\it ab initio} energies were also
calculated at the MRCI+Q/aug-cc-pV6Z and FCI/aug-cc-pV5Z level of
theory for off-grid geometries. They are also reported in Figure
S3 together with the RKHS energies evaluated at these
geometries. Again, the agreement between the electronic structure
calculations and the RKHS representation is good with $R^2$ value of
$(1-2\times10^{-7})$ for the MRCI+Q+LR PES.\\

\noindent
The equilibrium geometry of the FCI/aug-cc-pV5Z surface is a linear
He--H-H configuration ($r = 2.0749$ a$_{0}$ and $R=2.9712$ a$_{0}$),
with an energy of --2735.1 cm$^{-1}$ below the H$_{2}^{+}$
asymptote. This compares with the MRCI+Q+LR calculations for which
$r=2.0745$ a$_{0}$, $R=2.9713$ a$_{0}$ and depth --2736.2 cm$^{-1}$
and the earlier QCISD(T)/aug-cc-pvQz PES\cite{meu99:3418} ($r=2.0750$
a$_{0}$, $R=2.9720$ a$_{0}$ and depth --2717.0 cm$^{-1}$)
values. Hence, the structures of all PESs differ by less than 0.01
a$_0$ but the energetics varies over a range of $\sim 20$ cm$^{-1}$
whereby the dissociation energies for the two PESs from the present
work only differ by 1.1 cm$^{-1}$.\\

\subsection{Bound States}
The ground state energy of H$_2^+$($v=0,j=0)$--He computed from the
MRCI+Q+LR surface for $ortho-$H$_2^+$--He using DVR3D and CCVM are
--1795.1567 and --1795.3328 cm$^{-1}$, respectively. The same
energies, are obtained from the FCI PES as --1793.7632 and --1793.9067
cm$^{-1}$, using DVR3D and CCVM, respectively. These values are $\sim
40$ cm$^{-1}$ lower compared to those reported
previously\cite{meu99:3418} (--1754.269 cm$^{-1}$) on the QCISD(T)
PES. In Ref.\cite{meu99:3418} only 3 quadrature points along $r$ were
used for the 3D bound state calculations, which may not be sufficient
to fully converge the energies. The ground state energy is also
calculated in the present work following a time dependent wave packet
approach\cite{kon16:034303} on a 2D potential fixing $r$ at 2.0
a$_0$. These ground state energies are in fair agreement with previous
results (--1603 vs. --1593 cm$^{-1}$).\cite{meu99:3418}\\

\noindent
For {\it para-} H$_2^+$--He the ground state energies obtained from
the MRCI+Q+LR PES using DVR3D and CCVM are --1795.1575 and --1795.3352
cm$^{-1}$, respectively.  For the FCI surface the ground state
energies of {\it para-} H$_2^+$--He are calculated as --1793.7639 and
--1793.9091 cm$^{-1}$) using DVR3D and CCVM, respectively.  The
difference between DVR3D and CCVM is less than 0.17 cm$^{-1}$ for
both, $ortho$ and $para-$H$_2^+$--He. For the FCI+LR PES all the bound
states obtained from different methods for both, $ortho$ and
$para-$H$_2^+$--He are within 0.02 cm$^{-1}$ or less of the FCI PES
results. Hence, only the results obtained from the MRCI+Q+LR and FCI
PESs are reported.\\

\begin{figure}
\centering
\includegraphics[scale=0.5]{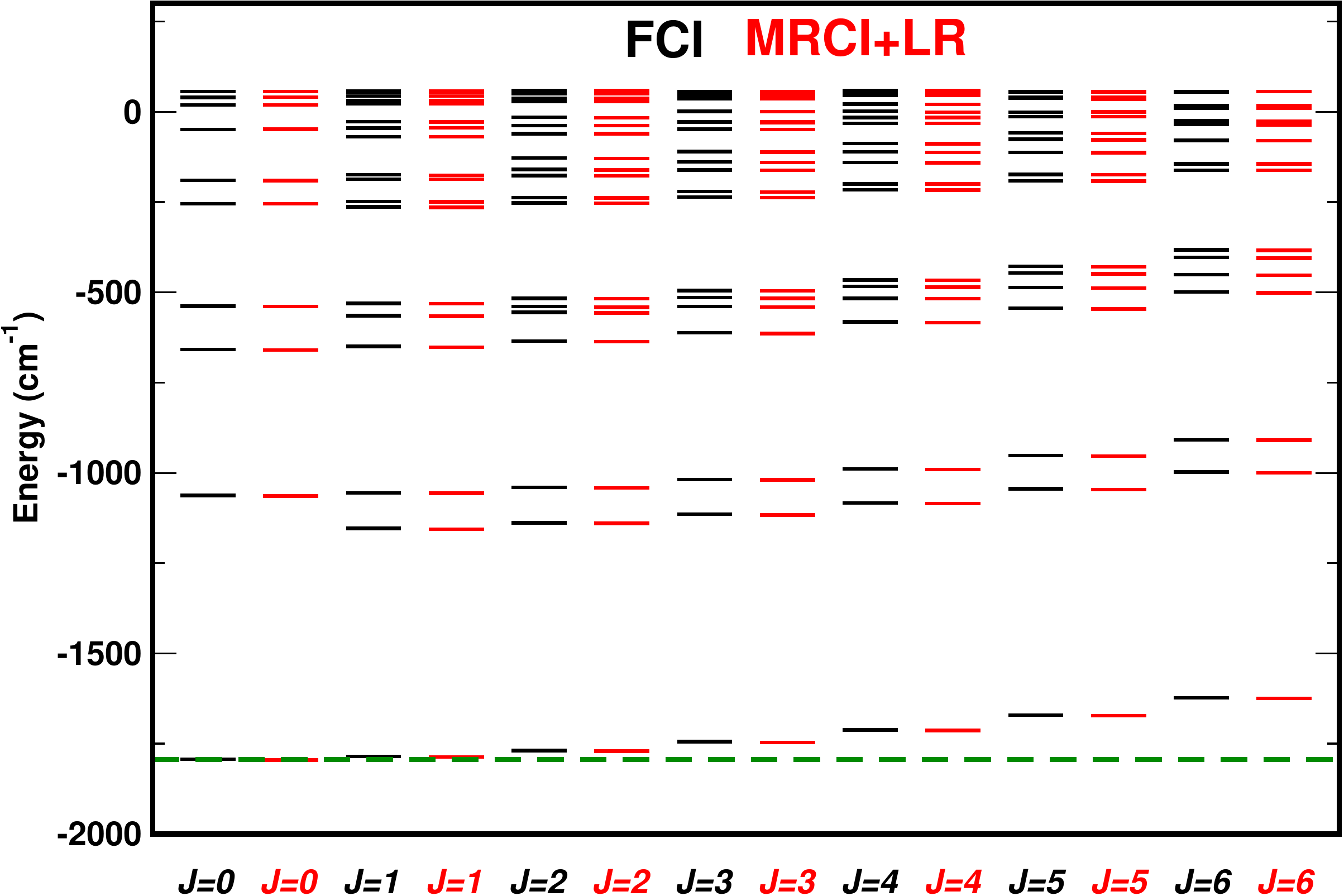}
\caption{Bound state energies for $J=0 \dots 6$ for {\it
    ortho-}H$_2^+$-He with $e$ symmetry). Results on the MRCI+Q+LR
  surface (black) and FCI (red) are shown. The ground state energy
  (--1793.7632 cm$^{−1}$) for the FCI/aug-cc-pv5Z surface is marked
  with a dashed line.}
\label{fig:fig4}
\end{figure}

\begin{figure}
    \centering
    \includegraphics[scale=0.50]{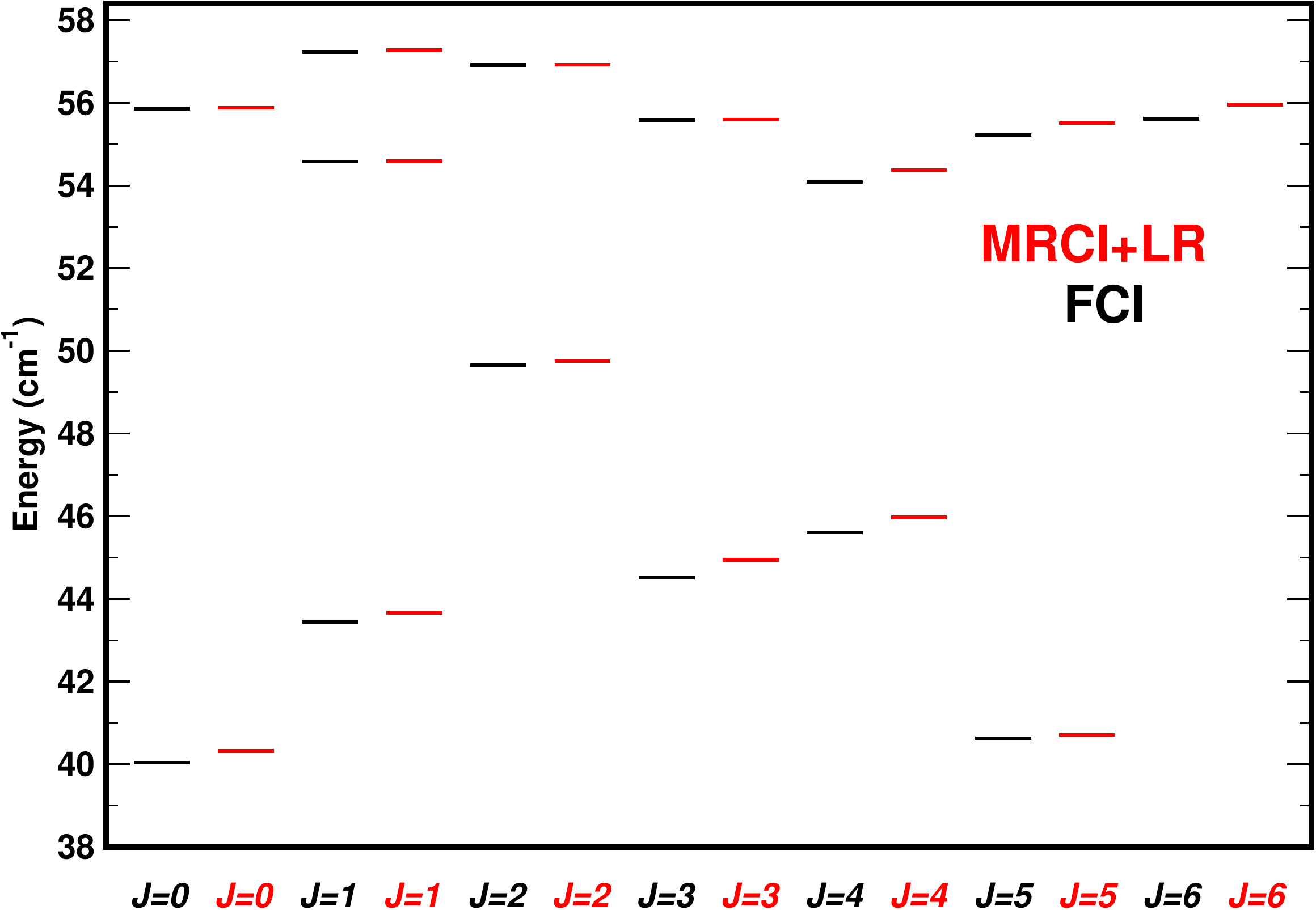}
    \caption{Bound state energies for ($J = 0 \dots 6$)
      (\textit{ortho}) with $e$ symmetry. With MRCI+Q+LR surface
      (black) and FCI (red). States within $\sim 20$ cm$^{-1}$ of the
      dissociation are reported.}
\label{fig:fig5}
\end{figure}

\noindent
A direct comparison for all and the near-dissociative (within 20
cm$^{-1}$ of dissociation) bound $e$ states for {\it
  ortho-}H$_2^+$--He and $J=0$ to $J=6$ from DVR3D calculations and
using the MRCI+Q+LR (red) and FCI (black) PESs is given in Figures
\ref{fig:fig4} and \ref{fig:fig5}. All states up to the dissociation
limit of $v=0, j=1$ state of H$_2^+$ are reported. The level pattern
for the two PESs is nearly identical. The distribution of the energy
difference $\Delta E$ between the MRCI+LR and the FCI PESs for
calculations with DVR3D or CCVM is given in Figure S5.\\

\noindent
The transitions that were probed by the microwave experiments lie
close to dissociation. Hence, a particular focus here is on accurately
computing these stationary states and to determine whether any
candidate transitions can be identified from using the MRCI+Q+LR and
the FCI PESs. A tentative assignment in particular for the 15.2 GHz
and 21.8 GHz transitions has been given previously based on
experiments using electric field dissociation.\cite{gam02:6072} They
were analyzed using an effective Hamiltonian. The 15.2 GHz transition
was assigned to a low-$N$ transition (in the terminology of
Ref.\cite{gam02:6072}, $N$ is the spin-free angular momentum which is
$J$ in the present work) with $\Delta N = 0$ with $N=3$ or $N=4$ in
{\it ortho-}H$_2^+$--He. In the following, $N$ is used when referring
to the analysis of the experiments\cite{gam02:6072} whereas $J$ is
used when discussing the present calculations. The fine and hyperfine
splittings due to coupling of electron and total nuclear spin, and
coupling of the resultant to the rotational angular momentum of the
nuclei are both less than 100 MHz, so are several orders of magnitude
smaller than the separations between rotational levels of the
complex. Thus, identification of $N$ with $J$ is a meaningful
approximation.\\

\noindent
For the 21.7 GHz transition on the other hand the analysis led to an
assignment involving $\Delta N = 1$ with $N = 11$ and $N' = 10$ in
{\it para-}H$_2^+$--He. While the analysis leading to a $\Delta N = 1$
transition involving {\it para-} H$_2^+$ is based on physical grounds,
that to a high-$N$ state involves fitting of the Zeeman pattern which
is more approximate. The selection rules for these transitions are $e
\leftrightarrow f$ for $\Delta J = 0$ and $e \leftrightarrow e$ or $f
\leftrightarrow f$ for $\Delta J = \pm 1$, respectively.\\

\begin{figure}
\centering
\includegraphics[scale=0.75]{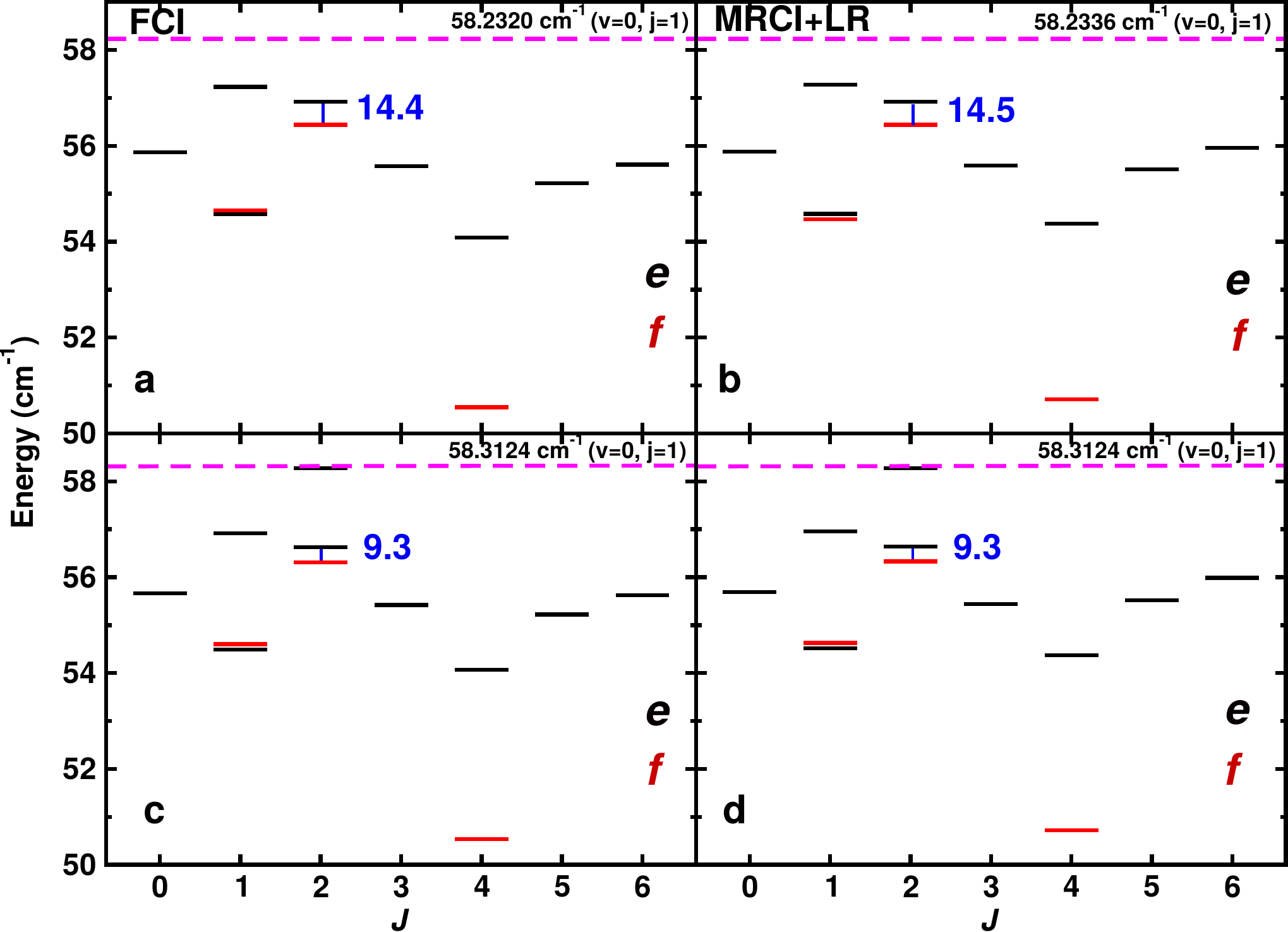}
\caption{Near dissociation {\it ortho-}H$_2^+$--He states (in
  cm$^{-1}$) and predicted transition frequencies in (GHz) using the
  MRCI+Q+LR (left, panels a and c) and FCI/aug-cc-pv5 (right, panels b
  and d) PESs. States with {\it e} (red) and {\it f} symmetry (black)
  are reported separately. Results from DVR3D and CCVM are in the top
  and bottom row, respectively. The 14.4(5) and 9.3 GHz parity doublet
  is a candidate for the 15.2 GHz line observed experimentally which
  had been assigned to a parity doublet with $\Delta N =
  0$.\cite{gam02:6072}}
\label{fig:fig6}
\end{figure}

\noindent
First, the near-dissociative states for {\it ortho-}H$_2^+$--He are
discussed. All near-dissociative states from the MRCI+Q+LR and FCI
PESs using the DVR3D and CCVM methods are reported in Figures
\ref{fig:fig6}a to d. All energies for $J=0$ to 6 are also reported in
Tables \ref{tab:mrci} and \ref{tab:fci}. There is one $e/f$ parity
doublet with $\Delta J = 0$ with a transition frequency between 10 and
18 GHz, involving the $J=2$ state for {\it ortho-}H$_2^+$--He. Using
DVR3D the transition frequency is 14.4(5) GHz whereas with the CCVM
code the transition is at 9.3 GHz. The parity doublet in both cases is
within 2 cm$^{-1}$ of dissociation which makes it a pair of
near-dissociative states. This is also confirmed by considering the
expectation value for the $R-$coordinate for the two states involved
which are $\langle R \rangle = 13.1$ a$_0$ for the $e$ state and
$\langle R \rangle = 12.7$ a$_0$ for the $f$ state which confirms
their long range character as suggested from the experiments.\\

\begin{figure}
\centering
\includegraphics[scale=0.75]{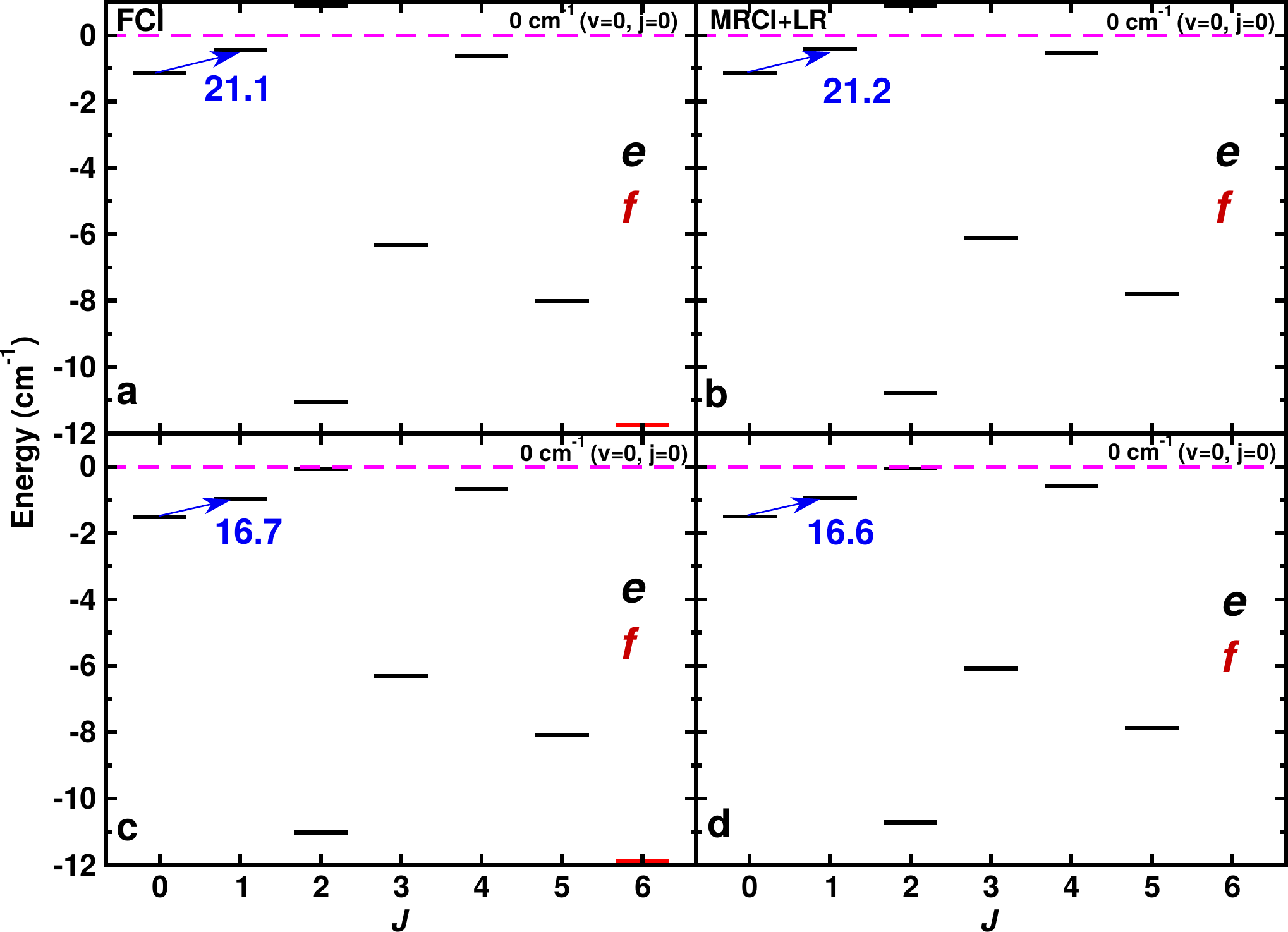}
\caption{Near dissociation {\it para-}H$_2^+$--He states (in
  cm$^{-1}$) and predicted transition frequencies in (GHz) using the
  MRCI+Q+LR (left, panels a and c) and FCI/aug-cc-pv5 (right, panels b
  and d) PESs. States with {\it e} (red) and {\it f} symmetry (black)
  are reported separately. Results from DVR3D and CCVM are in the top
  and bottom row, respectively.}
\label{fig:fig7}
\end{figure}

\noindent
Next, the near-dissociative states for {\it para-}H$_2^+$--He are
discussed for the two PESs and the two methods to compute bound
states, see Figures \ref{fig:fig7}a to d. The only near-dissociative
states involving either an $e/e$ or an $f/f$ transition with a
transition frequency around 20.1(2) and 16.7(6) GHz from DVR3D and
CCVM involves a $J=0$ and a $J=1$ state. The $\langle R \rangle =
15.0$ a$_0$ for the $J = 0, e$ state and $\langle R \rangle = 16.0$
a$_0$ for the $J = 1, f$ state, show the long range nature of the wave
functions. The potential candidate for the 21.8 GHz transition is not
found for the high $J$ states sufficiently close to dissociation to be
part of a suitable candidate transition.\\

\begin{table}[ht]
\caption{Near dissociation states calculated using MRCI surface with
  DVR3D and CCVM for He-H$_2^+$ in cm$^{-1}$. Zero is set to the
  energy of H$_2^+$($v=0,j=0$) state. The dissociation limit for
  $ortho$ is at 58.2336 cm$^{-1}$ and 58.3124 cm$^{-1}$ for the DVR3D
  and CCVM respectively, which corresponds to H$_2^+$($v=0,j=1$),
  whereas the dissociation limit for {\it para-} is at 0 cm$^{-1}$
  which corresponds to H$_2^+$($v=0,j=0$).}
\begin{tabular}{c|cc|cc|rr|rr}
\hline
\hline
    \multicolumn{5}{c}{\it ortho}& \multicolumn{4}{c}{\it para}\\
    \hline
    \multicolumn{3}{c}{$e$} & \multicolumn{2}{c}{$f$}& \multicolumn{2}{c}{$e$} & \multicolumn{2}{c}{$f$} \\
    \hline
    $J$ & CCVM   &  DVR3D  & CCVM  &  DVR3D  & CCVM    &    DVR3D &   CCVM  & DVR3D \\
   \hline
    0 & 39.896 & 40.321 &        &        & --15.641 & --15.719 \\
      & 55.689 & 55.876 &        &        &  --1.509 &  --1.137 \\
    1 & 43.199 & 43.668 & 54.629 & 54.464 & --13.965 & --14.035 \\
      & 54.514 & 54.582 &        &        &  --0.954 &  --0.430 \\
      & 56.959 & 57.273 &        &        &          &           \\
    2 & 49.345 & 49.755 & 56.329 & 56.439 & --10.712 & --10.766 \\
      & 56.639 & 56.923 &        &        &  --0.061 &    0.895 \\
      & 58.272 & 59.186 &        &        &          &           \\
    3 & 44.938 & 44.941 & 41.798 & 41.783 &  --6.082 & --6.109   \\
      & 55.443 & 55.588 &        &        &          &           \\
    4 & 46.022 & 45.971 & 50.718 & 50.712 &  --0.587 & --0.540   \\
      & 54.375 & 54.375 &        &        &          &           \\
    5 & 40.655 & 40.711 & 40.682 & 40.680 &  --7.877 & --7.799   \\
      & 55.518 & 55.507 &        &        &          &           \\
    6 & 55.985 & 55.959 &        &        &          &            & --13.897 & --13.746 \\
\hline
\hline
\end{tabular}
\label{tab:mrci}
\end{table}

\begin{table}[ht]
\caption{Near dissociation states calculated using FCI surface with
  DVR3D and CCVM for He-H$_2^+$ in cm$^{-1}$. Zero is set to the
  energy of H$_2^+$($v=0,j=0$) state. The dissociation limit for
  $ortho$ is at 58.2320 cm$^{-1}$ and 58.3124 cm$^{-1}$ for the DVR3D
  and CCVM respectively, which corresponds to H$_2^+$($v=0,j=1$),
  whereas the dissociation limit for $para$ is at 0 cm$^{-1}$ which
  corresponds to H$_2^+$($v=0,j=0$).}
  \begin{tabular}{c|cc|cc|rr|rr}
    \hline
    \hline
     \multicolumn{5}{c}{$ortho$}& \multicolumn{4}{c}{$para$}\\
    \hline
     \multicolumn{3}{c}{$e$} & \multicolumn{2}{c}{$f$}& \multicolumn{2}{c}{$e$} & \multicolumn{2}{c}{$f$} \\
    \hline
    \hline
      $J$ & CCVM   &  DVR3D & CCVM    & DVR3D  & CCVM      &  DVR3D \ & CCVM  & DVR3D\\
   \hline
    0 & 39.575 & 40.036 &         &        & --16.005 & --16.067\\
      & 55.662 & 55.860 &         &        &  --1.527 &  --1.150\\
    1 & 42.934 & 43.438 & 54.601  & 54.648 & --14.315 & --14.368\\
      & 54.494 & 54.575 &         &        &  --0.969 &  --0.448\\
      & 56.916 & 57.232 &         &        &          &         \\
    2 & 49.213 & 49.647 & 56.314  &  56.437 & --11.021 & --11.058\\
      & 56.623 & 56.916 &         &        &  --0.061 &    0.881\\
      & 58.271 & 59.162 &         &        &          &         \\
    3 & 44.497 & 44.513 & 41.473  & 41.473 & --6.311  &  --6.320\\
      & 55.422 & 55.577 &         &        &          &         \\
    4 & 45.652 & 45.611 & 50.538  & 50.547 & --0.680  &  --0.619\\
      & 54.067 & 54.084 &         &        &          &         \\
    5 & 40.565 & 40.626 & 40.854  & 40.860 & --8.092  &  --8.005\\
      & 55.219 & 55.220 &         &        &          & \\
    6 & 55.626 & 56.610 &         &        &          &  & --11.890  & --11.742\\
    \hline
    \hline
  \end{tabular}
  \label{tab:fci}
\end{table}

\section{Conclusions}
Two new PESs at the MRCI+Q+LR and FCI level of theory with large basis
sets and represented as a reproducing kernel have been used to
determine all bound and near-dissociative states for {\it ortho-} and
{\it para-}H$_2^+$--He. It is found that at both levels of theory the
bound states compare to within fractions of a wavenumber when
stationary states are determined from the same nuclear quantum
code. Moreover, the stationary states on one and the same PES
determined from two different quantum bound state codes (DVR3D and
CCVM) also agree closely, typically within less than fractions of one
cm$^{-1}$. This provides stringent benchmarks on potential transitions
that have been observed experimentally. One such assignment is for the
15.2 GHz transition which corresponds to {\it ortho-}H$_2^+$--He. The
transition found in the present work involves an $e/f$ parity doublet
with $J=2$. This compares with a tentative assignment to an $e/f$
parity doublet involving either a $J=3$ or $J=4$ state. For the 21.8
GHz transition which had been tentatively assigned to an $e/e$ or
$f/f$ transition in {\it para-}H$_2^{+}$--He the candidate,
near-dissociation states are $J=0$ and $J=1$, both of which are within
less than 2 cm$^{-1}$ of dissociation and the transition frequencies
range from 16 to 22 GHz. However, no high $J-$ candidate states
suitable for transition were found.\\

\noindent
The present work presents two high-accuracy, fully dimensional PESs
for H$_2^+$--He together with quantum bound state calculations that
provide potential assignments of experimentally characterized,
near-dissociation states. The results from both, MRCI+LR and full CI
PESs, using two different approaches for calculating the quantum bound
states are largely consistent. It will be interesting to use the
present PESs in future inelastic scattering calculations.\\

\section*{Conflicts of interest}
There are no conflicts of interest to declare.

\section*{Acknowledgments}
This work was supported by the Swiss National Science Foundation
through grants 200021-117810, the NCCR MUST and the AFOSR (to MM). The
authors thank Prof. J. Tennyson for exchange on the DVR3D program.

\bibliography{refs}

\begin{thebibliography}{40}%
\makeatletter
\providecommand \@ifxundefined [1]{%
 \@ifx{#1\undefined}
}%
\providecommand \@ifnum [1]{%
 \ifnum #1\expandafter \@firstoftwo
 \else \expandafter \@secondoftwo
 \fi
}%
\providecommand \@ifx [1]{%
 \ifx #1\expandafter \@firstoftwo
 \else \expandafter \@secondoftwo
 \fi
}%
\providecommand \natexlab [1]{#1}%
\providecommand \enquote  [1]{``#1''}%
\providecommand \bibnamefont  [1]{#1}%
\providecommand \bibfnamefont [1]{#1}%
\providecommand \citenamefont [1]{#1}%
\providecommand \href@noop [0]{\@secondoftwo}%
\providecommand \href [0]{\begingroup \@sanitize@url \@href}%
\providecommand \@href[1]{\@@startlink{#1}\@@href}%
\providecommand \@@href[1]{\endgroup#1\@@endlink}%
\providecommand \@sanitize@url [0]{\catcode `\\12\catcode `\$12\catcode
  `\&12\catcode `\#12\catcode `\^12\catcode `\_12\catcode `\%12\relax}%
\providecommand \@@startlink[1]{}%
\providecommand \@@endlink[0]{}%
\providecommand \url  [0]{\begingroup\@sanitize@url \@url }%
\providecommand \@url [1]{\endgroup\@href {#1}{\urlprefix }}%
\providecommand \urlprefix  [0]{URL }%
\providecommand \Eprint [0]{\href }%
\providecommand \doibase [0]{http://dx.doi.org/}%
\providecommand \selectlanguage [0]{\@gobble}%
\providecommand \bibinfo  [0]{\@secondoftwo}%
\providecommand \bibfield  [0]{\@secondoftwo}%
\providecommand \translation [1]{[#1]}%
\providecommand \BibitemOpen [0]{}%
\providecommand \bibitemStop [0]{}%
\providecommand \bibitemNoStop [0]{.\EOS\space}%
\providecommand \EOS [0]{\spacefactor3000\relax}%
\providecommand \BibitemShut  [1]{\csname bibitem#1\endcsname}%
\let\auto@bib@innerbib\@empty
\bibitem [{\citenamefont {Tielens}(2013)}]{tielens:2013}%
  \BibitemOpen
  \bibfield  {author} {\bibinfo {author} {\bibfnamefont {A.~G. G.~M.}\
  \bibnamefont {Tielens}},\ }\href@noop {} {\bibfield  {journal} {\bibinfo
  {journal} {Rev.\ Mod.\ Phys.}\ }\textbf {\bibinfo {volume} {85}},\ \bibinfo
  {pages} {1021} (\bibinfo {year} {2013})}\BibitemShut {NoStop}%
\bibitem [{\citenamefont {Li}, \citenamefont {Jiang},\ and\ \citenamefont
  {Hao}(2015)}]{hao:2015}%
  \BibitemOpen
  \bibfield  {author} {\bibinfo {author} {\bibfnamefont {Q.}~\bibnamefont
  {Li}}, \bibinfo {author} {\bibfnamefont {J.}~\bibnamefont {Jiang}}, \ and\
  \bibinfo {author} {\bibfnamefont {J.}~\bibnamefont {Hao}},\ }\href@noop {}
  {\bibfield  {journal} {\bibinfo  {journal} {Kona Powder Part. J.}\ }\textbf
  {\bibinfo {volume} {32}},\ \bibinfo {pages} {57} (\bibinfo {year}
  {2015})}\BibitemShut {NoStop}%
\bibitem [{\citenamefont {Guesten}\ \emph {et~al.}(2019)\citenamefont
  {Guesten}, \citenamefont {Wiesemeyer}, \citenamefont {Neufeld}, \citenamefont
  {Menten}, \citenamefont {Graf}, \citenamefont {Jacobs}, \citenamefont
  {Klein}, \citenamefont {Ricken}, \citenamefont {Risacher},\ and\
  \citenamefont {Stutzki}}]{stutzki:2019}%
  \BibitemOpen
  \bibfield  {author} {\bibinfo {author} {\bibfnamefont {R.}~\bibnamefont
  {Guesten}}, \bibinfo {author} {\bibfnamefont {H.}~\bibnamefont {Wiesemeyer}},
  \bibinfo {author} {\bibfnamefont {D.}~\bibnamefont {Neufeld}}, \bibinfo
  {author} {\bibfnamefont {K.~M.}\ \bibnamefont {Menten}}, \bibinfo {author}
  {\bibfnamefont {U.~U.}\ \bibnamefont {Graf}}, \bibinfo {author}
  {\bibfnamefont {K.}~\bibnamefont {Jacobs}}, \bibinfo {author} {\bibfnamefont
  {B.}~\bibnamefont {Klein}}, \bibinfo {author} {\bibfnamefont
  {O.}~\bibnamefont {Ricken}}, \bibinfo {author} {\bibfnamefont
  {C.}~\bibnamefont {Risacher}}, \ and\ \bibinfo {author} {\bibfnamefont
  {J.}~\bibnamefont {Stutzki}},\ }\href@noop {} {\bibfield  {journal} {\bibinfo
   {journal} {Nature}\ }\textbf {\bibinfo {volume} {568}},\ \bibinfo {pages}
  {357} (\bibinfo {year} {2019})}\BibitemShut {NoStop}%
\bibitem [{\citenamefont {Zygelman}, \citenamefont {Stancil},\ and\
  \citenamefont {Dalgarno}(1998)}]{zyg98:151}%
  \BibitemOpen
  \bibfield  {author} {\bibinfo {author} {\bibfnamefont {B.}~\bibnamefont
  {Zygelman}}, \bibinfo {author} {\bibfnamefont {P.~C.}\ \bibnamefont
  {Stancil}}, \ and\ \bibinfo {author} {\bibfnamefont {A.}~\bibnamefont
  {Dalgarno}},\ }\href@noop {} {\bibfield  {journal} {\bibinfo  {journal}
  {Astrophys.\ J.}\ }\textbf {\bibinfo {volume} {508}},\ \bibinfo {pages} {151}
  (\bibinfo {year} {1998})}\BibitemShut {NoStop}%
\bibitem [{\citenamefont {Black}(2012)}]{black:2012}%
  \BibitemOpen
  \bibfield  {author} {\bibinfo {author} {\bibfnamefont {J.~H.}\ \bibnamefont
  {Black}},\ }\href@noop {} {\bibfield  {journal} {\bibinfo  {journal} {Philos.
  Trans. Royal Soc. A}\ }\textbf {\bibinfo {volume} {370}},\ \bibinfo {pages}
  {5130} (\bibinfo {year} {2012})}\BibitemShut {NoStop}%
\bibitem [{\citenamefont {Lepp}, \citenamefont {Stancil},\ and\ \citenamefont
  {Dalgarno}(2002)}]{lep02:R57}%
  \BibitemOpen
  \bibfield  {author} {\bibinfo {author} {\bibfnamefont {S.}~\bibnamefont
  {Lepp}}, \bibinfo {author} {\bibfnamefont {P.~C.}\ \bibnamefont {Stancil}}, \
  and\ \bibinfo {author} {\bibfnamefont {A.}~\bibnamefont {Dalgarno}},\
  }\href@noop {} {\bibfield  {journal} {\bibinfo  {journal}
  {J.~Phys.~B:~At.~Mol.~Opt.~Phys.}\ }\textbf {\bibinfo {volume} {35}},\
  \bibinfo {pages} {R57} (\bibinfo {year} {2002})}\BibitemShut {NoStop}%
\bibitem [{\citenamefont {Vera}\ \emph {et~al.}(2017)\citenamefont {Vera},
  \citenamefont {Gianturco}, \citenamefont {Wester}, \citenamefont {da~Silva},
  \citenamefont {Dulieu},\ and\ \citenamefont {Schiller}}]{schiller:2017}%
  \BibitemOpen
  \bibfield  {author} {\bibinfo {author} {\bibfnamefont {M.~H.}\ \bibnamefont
  {Vera}}, \bibinfo {author} {\bibfnamefont {F.~A.}\ \bibnamefont {Gianturco}},
  \bibinfo {author} {\bibfnamefont {R.}~\bibnamefont {Wester}}, \bibinfo
  {author} {\bibfnamefont {H.}~\bibnamefont {da~Silva}, \bibfnamefont {Jr.}},
  \bibinfo {author} {\bibfnamefont {O.}~\bibnamefont {Dulieu}}, \ and\ \bibinfo
  {author} {\bibfnamefont {S.}~\bibnamefont {Schiller}},\ }\href@noop {}
  {\bibfield  {journal} {\bibinfo  {journal} {J.~Chem.\ Phys.}\ }\textbf
  {\bibinfo {volume} {{146}}} (\bibinfo {year} {{2017}})}\BibitemShut {NoStop}%
\bibitem [{\citenamefont {Koelemeij}\ \emph {et~al.}(2007)\citenamefont
  {Koelemeij}, \citenamefont {Roth}, \citenamefont {Wicht}, \citenamefont
  {Ernsting},\ and\ \citenamefont {Schiller}}]{schiller:2007}%
  \BibitemOpen
  \bibfield  {author} {\bibinfo {author} {\bibfnamefont {J.~C.~J.}\
  \bibnamefont {Koelemeij}}, \bibinfo {author} {\bibfnamefont {B.}~\bibnamefont
  {Roth}}, \bibinfo {author} {\bibfnamefont {A.}~\bibnamefont {Wicht}},
  \bibinfo {author} {\bibfnamefont {I.}~\bibnamefont {Ernsting}}, \ and\
  \bibinfo {author} {\bibfnamefont {S.}~\bibnamefont {Schiller}},\ }\href@noop
  {} {\bibfield  {journal} {\bibinfo  {journal} {Phys.\ Rev.\ Lett.}\ }\textbf
  {\bibinfo {volume} {{98}}},\ \bibinfo {pages} {{173002}} (\bibinfo {year}
  {{2007}})}\BibitemShut {NoStop}%
\bibitem [{\citenamefont {Schiller}, \citenamefont {Bakalov},\ and\
  \citenamefont {Korobov}(2014)}]{schiller:2014}%
  \BibitemOpen
  \bibfield  {author} {\bibinfo {author} {\bibfnamefont {S.}~\bibnamefont
  {Schiller}}, \bibinfo {author} {\bibfnamefont {D.}~\bibnamefont {Bakalov}}, \
  and\ \bibinfo {author} {\bibfnamefont {V.~I.}\ \bibnamefont {Korobov}},\
  }\href@noop {} {\bibfield  {journal} {\bibinfo  {journal} {Phys.\ Rev.\
  Lett.}\ }\textbf {\bibinfo {volume} {{113}}},\ \bibinfo {pages} {{023004}}
  (\bibinfo {year} {{2014}})}\BibitemShut {NoStop}%
\bibitem [{\citenamefont {Klein}\ \emph {et~al.}(2017)\citenamefont {Klein},
  \citenamefont {Shagam}, \citenamefont {Skomorowski}, \citenamefont
  {Zuchowski}, \citenamefont {Pawlak}, \citenamefont {Janssen}, \citenamefont
  {Moiseyev}, \citenamefont {van~de Meerakker}, \citenamefont {van~der Avoird},
  \citenamefont {Koch},\ and\ \citenamefont {Narevicius}}]{klein:2017}%
  \BibitemOpen
  \bibfield  {author} {\bibinfo {author} {\bibfnamefont {A.}~\bibnamefont
  {Klein}}, \bibinfo {author} {\bibfnamefont {Y.}~\bibnamefont {Shagam}},
  \bibinfo {author} {\bibfnamefont {W.}~\bibnamefont {Skomorowski}}, \bibinfo
  {author} {\bibfnamefont {P.~S.}\ \bibnamefont {Zuchowski}}, \bibinfo {author}
  {\bibfnamefont {M.}~\bibnamefont {Pawlak}}, \bibinfo {author} {\bibfnamefont
  {L.~M.~C.}\ \bibnamefont {Janssen}}, \bibinfo {author} {\bibfnamefont
  {N.}~\bibnamefont {Moiseyev}}, \bibinfo {author} {\bibfnamefont {S.~Y.~T.}\
  \bibnamefont {van~de Meerakker}}, \bibinfo {author} {\bibfnamefont
  {A.}~\bibnamefont {van~der Avoird}}, \bibinfo {author} {\bibfnamefont
  {C.~P.}\ \bibnamefont {Koch}}, \ and\ \bibinfo {author} {\bibfnamefont
  {E.}~\bibnamefont {Narevicius}},\ }\href@noop {} {\bibfield  {journal}
  {\bibinfo  {journal} {J.~Chem.\ Phys.}\ }\textbf {\bibinfo {volume} {13}},\
  \bibinfo {pages} {35} (\bibinfo {year} {2017})}\BibitemShut {NoStop}%
\bibitem [{\citenamefont {Carrington}\ \emph {et~al.}(1996)\citenamefont
  {Carrington}, \citenamefont {Gammie}, \citenamefont {Shaw}, \citenamefont
  {Taylor},\ and\ \citenamefont {Hutson}}]{car96:395}%
  \BibitemOpen
  \bibfield  {author} {\bibinfo {author} {\bibfnamefont {A.}~\bibnamefont
  {Carrington}}, \bibinfo {author} {\bibfnamefont {D.~I.}\ \bibnamefont
  {Gammie}}, \bibinfo {author} {\bibfnamefont {A.~M.}\ \bibnamefont {Shaw}},
  \bibinfo {author} {\bibfnamefont {S.~M.}\ \bibnamefont {Taylor}}, \ and\
  \bibinfo {author} {\bibfnamefont {J.~M.}\ \bibnamefont {Hutson}},\
  }\href@noop {} {\bibfield  {journal} {\bibinfo  {journal} {Chem. Phys.
  Lett.}\ }\textbf {\bibinfo {volume} {260}},\ \bibinfo {pages} {395} (\bibinfo
  {year} {1996})}\BibitemShut {NoStop}%
\bibitem [{\citenamefont {Gammie}, \citenamefont {Page},\ and\ \citenamefont
  {Shaw}(2002)}]{gam02:6072}%
  \BibitemOpen
  \bibfield  {author} {\bibinfo {author} {\bibfnamefont {D.~I.}\ \bibnamefont
  {Gammie}}, \bibinfo {author} {\bibfnamefont {J.~C.}\ \bibnamefont {Page}}, \
  and\ \bibinfo {author} {\bibfnamefont {A.~M.}\ \bibnamefont {Shaw}},\
  }\href@noop {} {\bibfield  {journal} {\bibinfo  {journal} {J.~Chem.\ Phys.}\
  }\textbf {\bibinfo {volume} {116}},\ \bibinfo {pages} {6072} (\bibinfo {year}
  {2002})}\BibitemShut {NoStop}%
\bibitem [{\citenamefont {Joseph}\ and\ \citenamefont
  {Sathyamurthy}(1987)}]{jos87:704}%
  \BibitemOpen
  \bibfield  {author} {\bibinfo {author} {\bibfnamefont {T.}~\bibnamefont
  {Joseph}}\ and\ \bibinfo {author} {\bibfnamefont {N.}~\bibnamefont
  {Sathyamurthy}},\ }\href@noop {} {\bibfield  {journal} {\bibinfo  {journal}
  {J. Chem. Phys.}\ }\textbf {\bibinfo {volume} {86}},\ \bibinfo {pages} {704}
  (\bibinfo {year} {1987})}\BibitemShut {NoStop}%
\bibitem [{\citenamefont {Falcetta}\ and\ \citenamefont
  {Siska}(1999)}]{fal99:117}%
  \BibitemOpen
  \bibfield  {author} {\bibinfo {author} {\bibfnamefont {M.~F.}\ \bibnamefont
  {Falcetta}}\ and\ \bibinfo {author} {\bibfnamefont {P.~E.}\ \bibnamefont
  {Siska}},\ }\href@noop {} {\bibfield  {journal} {\bibinfo  {journal} {Mol.
  Phys}\ }\textbf {\bibinfo {volume} {97}},\ \bibinfo {pages} {117} (\bibinfo
  {year} {1999})}\BibitemShut {NoStop}%
\bibitem [{\citenamefont {Meuwly}\ and\ \citenamefont
  {Hutson}(1999)}]{meu99:3418}%
  \BibitemOpen
  \bibfield  {author} {\bibinfo {author} {\bibfnamefont {M.}~\bibnamefont
  {Meuwly}}\ and\ \bibinfo {author} {\bibfnamefont {J.~M.}\ \bibnamefont
  {Hutson}},\ }\href@noop {} {\bibfield  {journal} {\bibinfo  {journal}
  {J.~Chem.\ Phys.}\ }\textbf {\bibinfo {volume} {110}},\ \bibinfo {pages}
  {3418} (\bibinfo {year} {1999})}\BibitemShut {NoStop}%
\bibitem [{\citenamefont {Palmieri}\ \emph {et~al.}(2000)\citenamefont
  {Palmieri}, \citenamefont {Puzzarini}, \citenamefont {Aquilanti},
  \citenamefont {Capecchi}, \citenamefont {Cavalli}, \citenamefont {de~Fazio},
  \citenamefont {Aguilar}, \citenamefont {Gim\'{e}nez},\ and\ \citenamefont
  {Lucas}}]{pal00:1835}%
  \BibitemOpen
  \bibfield  {author} {\bibinfo {author} {\bibfnamefont {P.}~\bibnamefont
  {Palmieri}}, \bibinfo {author} {\bibfnamefont {C.}~\bibnamefont {Puzzarini}},
  \bibinfo {author} {\bibfnamefont {V.}~\bibnamefont {Aquilanti}}, \bibinfo
  {author} {\bibfnamefont {G.}~\bibnamefont {Capecchi}}, \bibinfo {author}
  {\bibfnamefont {S.}~\bibnamefont {Cavalli}}, \bibinfo {author} {\bibfnamefont
  {D.}~\bibnamefont {de~Fazio}}, \bibinfo {author} {\bibfnamefont
  {A.}~\bibnamefont {Aguilar}}, \bibinfo {author} {\bibfnamefont
  {X.}~\bibnamefont {Gim\'{e}nez}}, \ and\ \bibinfo {author} {\bibfnamefont
  {J.~M.}\ \bibnamefont {Lucas}},\ }\href@noop {} {\bibfield  {journal}
  {\bibinfo  {journal} {Mol. Phys.}\ }\textbf {\bibinfo {volume} {98}},\
  \bibinfo {pages} {1835} (\bibinfo {year} {2000})}\BibitemShut {NoStop}%
\bibitem [{\citenamefont {Ramachandran}\ \emph {et~al.}(2009)\citenamefont
  {Ramachandran}, \citenamefont {Fazio}, \citenamefont {Cavalli}, \citenamefont
  {Tarantelli},\ and\ \citenamefont {Aquilanti}}]{ram09:26}%
  \BibitemOpen
  \bibfield  {author} {\bibinfo {author} {\bibfnamefont {C.}~\bibnamefont
  {Ramachandran}}, \bibinfo {author} {\bibfnamefont {D.~D.}\ \bibnamefont
  {Fazio}}, \bibinfo {author} {\bibfnamefont {S.}~\bibnamefont {Cavalli}},
  \bibinfo {author} {\bibfnamefont {F.}~\bibnamefont {Tarantelli}}, \ and\
  \bibinfo {author} {\bibfnamefont {V.}~\bibnamefont {Aquilanti}},\ }\href@noop
  {} {\bibfield  {journal} {\bibinfo  {journal} {Chem.\ Phys.\ Lett.}\ }\textbf
  {\bibinfo {volume} {469}},\ \bibinfo {pages} {26 } (\bibinfo {year}
  {2009})}\BibitemShut {NoStop}%
\bibitem [{\citenamefont {de~Fazio}\ \emph {et~al.}(2012)\citenamefont
  {de~Fazio}, \citenamefont {de~Castro-Vitores}, \citenamefont {Aguado},
  \citenamefont {Aquilanti},\ and\ \citenamefont {Cavalli}}]{faz12:244306}%
  \BibitemOpen
  \bibfield  {author} {\bibinfo {author} {\bibfnamefont {D.}~\bibnamefont
  {de~Fazio}}, \bibinfo {author} {\bibfnamefont {M.}~\bibnamefont
  {de~Castro-Vitores}}, \bibinfo {author} {\bibfnamefont {A.}~\bibnamefont
  {Aguado}}, \bibinfo {author} {\bibfnamefont {V.}~\bibnamefont {Aquilanti}}, \
  and\ \bibinfo {author} {\bibfnamefont {S.}~\bibnamefont {Cavalli}},\
  }\href@noop {} {\bibfield  {journal} {\bibinfo  {journal} {J.~Chem.\ Phys.}\
  }\textbf {\bibinfo {volume} {137}},\ \bibinfo {pages} {244306} (\bibinfo
  {year} {2012})}\BibitemShut {NoStop}%
\bibitem [{\citenamefont {Werner}\ and\ \citenamefont
  {Knowles}(1988)}]{wer88:5803}%
  \BibitemOpen
  \bibfield  {author} {\bibinfo {author} {\bibfnamefont {H.}~\bibnamefont
  {Werner}}\ and\ \bibinfo {author} {\bibfnamefont {P.~J.}\ \bibnamefont
  {Knowles}},\ }\href@noop {} {\bibfield  {journal} {\bibinfo  {journal}
  {J.~Chem.\ Phys.}\ }\textbf {\bibinfo {volume} {89}},\ \bibinfo {pages}
  {5803} (\bibinfo {year} {1988})}\BibitemShut {NoStop}%
\bibitem [{\citenamefont {Knowles}\ and\ \citenamefont
  {Werner}(1988)}]{kno88:514}%
  \BibitemOpen
  \bibfield  {author} {\bibinfo {author} {\bibfnamefont {P.~J.}\ \bibnamefont
  {Knowles}}\ and\ \bibinfo {author} {\bibfnamefont {H.-J.}\ \bibnamefont
  {Werner}},\ }\href@noop {} {\bibfield  {journal} {\bibinfo  {journal} {Chem.\
  Phys.\ Lett.}\ }\textbf {\bibinfo {volume} {145}},\ \bibinfo {pages} {514 }
  (\bibinfo {year} {1988})}\BibitemShut {NoStop}%
\bibitem [{\citenamefont {Wilson}, \citenamefont {van Mourik},\ and\
  \citenamefont {Dunning}(1996)}]{wil96:339}%
  \BibitemOpen
  \bibfield  {author} {\bibinfo {author} {\bibfnamefont {A.~K.}\ \bibnamefont
  {Wilson}}, \bibinfo {author} {\bibfnamefont {T.}~\bibnamefont {van Mourik}},
  \ and\ \bibinfo {author} {\bibfnamefont {T.~H.}\ \bibnamefont {Dunning}},\
  }\href@noop {} {\bibfield  {journal} {\bibinfo  {journal} {J. Mol. Struct.
  (Theochem)}\ }\textbf {\bibinfo {volume} {388}},\ \bibinfo {pages} {339 }
  (\bibinfo {year} {1996})}\BibitemShut {NoStop}%
\bibitem [{\citenamefont {Knowles}\ and\ \citenamefont
  {Handy}(1984)}]{kno84:315}%
  \BibitemOpen
  \bibfield  {author} {\bibinfo {author} {\bibfnamefont {P.}~\bibnamefont
  {Knowles}}\ and\ \bibinfo {author} {\bibfnamefont {N.}~\bibnamefont
  {Handy}},\ }\href@noop {} {\bibfield  {journal} {\bibinfo  {journal} {Chem.\
  Phys.\ Lett.}\ }\textbf {\bibinfo {volume} {111}},\ \bibinfo {pages} {315 }
  (\bibinfo {year} {1984})}\BibitemShut {NoStop}%
\bibitem [{\citenamefont {Knowles}\ and\ \citenamefont
  {Handy}(1989)}]{kno89:75}%
  \BibitemOpen
  \bibfield  {author} {\bibinfo {author} {\bibfnamefont {P.~J.}\ \bibnamefont
  {Knowles}}\ and\ \bibinfo {author} {\bibfnamefont {N.~C.}\ \bibnamefont
  {Handy}},\ }\href@noop {} {\bibfield  {journal} {\bibinfo  {journal}
  {Comput.\ Phys.\ Commun.}\ }\textbf {\bibinfo {volume} {54}},\ \bibinfo
  {pages} {75 } (\bibinfo {year} {1989})}\BibitemShut {NoStop}%
\bibitem [{\citenamefont {Dunning}(1989)}]{dun89:1007}%
  \BibitemOpen
  \bibfield  {author} {\bibinfo {author} {\bibfnamefont {T.~H.}\ \bibnamefont
  {Dunning}},\ }\href@noop {} {\bibfield  {journal} {\bibinfo  {journal}
  {J.~Chem.\ Phys.}\ }\textbf {\bibinfo {volume} {90}},\ \bibinfo {pages}
  {1007} (\bibinfo {year} {1989})}\BibitemShut {NoStop}%
\bibitem [{\citenamefont {Woon}\ and\ \citenamefont
  {Dunning}(1994)}]{woo94:2975}%
  \BibitemOpen
  \bibfield  {author} {\bibinfo {author} {\bibfnamefont {D.~E.}\ \bibnamefont
  {Woon}}\ and\ \bibinfo {author} {\bibfnamefont {T.~H.}\ \bibnamefont
  {Dunning}},\ }\href@noop {} {\bibfield  {journal} {\bibinfo  {journal}
  {J.~Chem.\ Phys.}\ }\textbf {\bibinfo {volume} {100}},\ \bibinfo {pages}
  {2975} (\bibinfo {year} {1994})}\BibitemShut {NoStop}%
\bibitem [{\citenamefont {Werner}\ and\ \citenamefont
  {Knowles}(1985)}]{wen85:5053}%
  \BibitemOpen
  \bibfield  {author} {\bibinfo {author} {\bibfnamefont {H.}~\bibnamefont
  {Werner}}\ and\ \bibinfo {author} {\bibfnamefont {P.~J.}\ \bibnamefont
  {Knowles}},\ }\href@noop {} {\bibfield  {journal} {\bibinfo  {journal}
  {J.~Chem.\ Phys.}\ }\textbf {\bibinfo {volume} {82}},\ \bibinfo {pages}
  {5053} (\bibinfo {year} {1985})}\BibitemShut {NoStop}%
\bibitem [{\citenamefont {Knowles}\ and\ \citenamefont
  {Werner}(1985)}]{kno85:259}%
  \BibitemOpen
  \bibfield  {author} {\bibinfo {author} {\bibfnamefont {P.~J.}\ \bibnamefont
  {Knowles}}\ and\ \bibinfo {author} {\bibfnamefont {H.-J.}\ \bibnamefont
  {Werner}},\ }\href@noop {} {\bibfield  {journal} {\bibinfo  {journal} {Chem.\
  Phys.\ Lett.}\ }\textbf {\bibinfo {volume} {115}},\ \bibinfo {pages} {259 }
  (\bibinfo {year} {1985})}\BibitemShut {NoStop}%
\bibitem [{\citenamefont {Werner}\ and\ \citenamefont
  {Meyer}(1980)}]{wer80:2342}%
  \BibitemOpen
  \bibfield  {author} {\bibinfo {author} {\bibfnamefont {H.}~\bibnamefont
  {Werner}}\ and\ \bibinfo {author} {\bibfnamefont {W.}~\bibnamefont {Meyer}},\
  }\href@noop {} {\bibfield  {journal} {\bibinfo  {journal} {J.~Chem.\ Phys.}\
  }\textbf {\bibinfo {volume} {73}},\ \bibinfo {pages} {2342} (\bibinfo {year}
  {1980})}\BibitemShut {NoStop}%
\bibitem [{\citenamefont {Werner}\ \emph {et~al.}(2012)\citenamefont {Werner},
  \citenamefont {Knowles}, \citenamefont {Knizia}, \citenamefont {Manby},\ and\
  \citenamefont {Sch\"{u}tz}}]{molpro}%
  \BibitemOpen
  \bibfield  {author} {\bibinfo {author} {\bibfnamefont {H.-J.}\ \bibnamefont
  {Werner}}, \bibinfo {author} {\bibfnamefont {P.~J.}\ \bibnamefont {Knowles}},
  \bibinfo {author} {\bibfnamefont {G.}~\bibnamefont {Knizia}}, \bibinfo
  {author} {\bibfnamefont {F.~R.}\ \bibnamefont {Manby}}, \ and\ \bibinfo
  {author} {\bibfnamefont {M.}~\bibnamefont {Sch\"{u}tz}},\ }\href@noop {}
  {\bibfield  {journal} {\bibinfo  {journal} {Wiley Interdiscip. Rev.: Comput.
  Mol. Sci.}\ }\textbf {\bibinfo {volume} {2}},\ \bibinfo {pages} {242}
  (\bibinfo {year} {2012})}\BibitemShut {NoStop}%
\bibitem [{\citenamefont {Varandas}(1988)}]{var88:255}%
  \BibitemOpen
  \bibfield  {author} {\bibinfo {author} {\bibfnamefont {A.~J.~C.}\
  \bibnamefont {Varandas}},\ }\href@noop {} {\bibfield  {journal} {\bibinfo
  {journal} {Adv. Chem. Phys.}\ }\textbf {\bibinfo {volume} {74}},\ \bibinfo
  {pages} {255} (\bibinfo {year} {1988})}\BibitemShut {NoStop}%
\bibitem [{\citenamefont {Aguado}\ and\ \citenamefont
  {Paniagua}(1992)}]{agu92:1265}%
  \BibitemOpen
  \bibfield  {author} {\bibinfo {author} {\bibfnamefont {A.}~\bibnamefont
  {Aguado}}\ and\ \bibinfo {author} {\bibfnamefont {M.}~\bibnamefont
  {Paniagua}},\ }\href@noop {} {\bibfield  {journal} {\bibinfo  {journal} {J.
  Chem. Phys.}\ }\textbf {\bibinfo {volume} {96}},\ \bibinfo {pages} {1265}
  (\bibinfo {year} {1992})}\BibitemShut {NoStop}%
\bibitem [{\citenamefont {Bishop}\ and\ \citenamefont
  {Pipin}(1995)}]{bis95:15}%
  \BibitemOpen
  \bibfield  {author} {\bibinfo {author} {\bibfnamefont {D.~M.}\ \bibnamefont
  {Bishop}}\ and\ \bibinfo {author} {\bibfnamefont {J.}~\bibnamefont {Pipin}},\
  }\href@noop {} {\bibfield  {journal} {\bibinfo  {journal} {Chem.\ Phys.\
  Lett.}\ }\textbf {\bibinfo {volume} {236}},\ \bibinfo {pages} {15} (\bibinfo
  {year} {1995})}\BibitemShut {NoStop}%
\bibitem [{\citenamefont {Velilla}\ \emph {et~al.}(2008)\citenamefont
  {Velilla}, \citenamefont {Lepetit}, \citenamefont {Aguado}, \citenamefont
  {Beswick},\ and\ \citenamefont {Paniagua}}]{vel08:084307}%
  \BibitemOpen
  \bibfield  {author} {\bibinfo {author} {\bibfnamefont {L.}~\bibnamefont
  {Velilla}}, \bibinfo {author} {\bibfnamefont {B.}~\bibnamefont {Lepetit}},
  \bibinfo {author} {\bibfnamefont {A.}~\bibnamefont {Aguado}}, \bibinfo
  {author} {\bibfnamefont {A.}~\bibnamefont {Beswick}}, \ and\ \bibinfo
  {author} {\bibfnamefont {M.}~\bibnamefont {Paniagua}},\ }\href@noop {}
  {\bibfield  {journal} {\bibinfo  {journal} {J.~Chem.\ Phys.}\ }\textbf
  {\bibinfo {volume} {129}},\ \bibinfo {pages} {084307} (\bibinfo {year}
  {2008})}\BibitemShut {NoStop}%
\bibitem [{\citenamefont {Press}\ \emph {et~al.}(1992)\citenamefont {Press},
  \citenamefont {Teukolsky}, \citenamefont {Vetterling},\ and\ \citenamefont
  {Flannery}}]{pre92}%
  \BibitemOpen
  \bibfield  {author} {\bibinfo {author} {\bibfnamefont {W.~H.}\ \bibnamefont
  {Press}}, \bibinfo {author} {\bibfnamefont {S.~A.}\ \bibnamefont
  {Teukolsky}}, \bibinfo {author} {\bibfnamefont {W.~T.}\ \bibnamefont
  {Vetterling}}, \ and\ \bibinfo {author} {\bibfnamefont {B.~P.}\ \bibnamefont
  {Flannery}},\ }\href@noop {} {\emph {\bibinfo {title} {{Numerical Recipes in
  Fortran 77}}}}\ (\bibinfo  {publisher} {Cambridge University Press, New
  York},\ \bibinfo {year} {1992})\BibitemShut {NoStop}%
\bibitem [{\citenamefont {Ho}\ and\ \citenamefont {Rabitz}(1996)}]{ho96:2584}%
  \BibitemOpen
  \bibfield  {author} {\bibinfo {author} {\bibfnamefont {T.-S.}\ \bibnamefont
  {Ho}}\ and\ \bibinfo {author} {\bibfnamefont {H.}~\bibnamefont {Rabitz}},\
  }\href@noop {} {\bibfield  {journal} {\bibinfo  {journal} {J.~Chem.\ Phys.}\
  }\textbf {\bibinfo {volume} {104}},\ \bibinfo {pages} {2584} (\bibinfo {year}
  {1996})}\BibitemShut {NoStop}%
\bibitem [{\citenamefont {Unke}\ and\ \citenamefont
  {Meuwly}(2017)}]{unk17:1923}%
  \BibitemOpen
  \bibfield  {author} {\bibinfo {author} {\bibfnamefont {O.~T.}\ \bibnamefont
  {Unke}}\ and\ \bibinfo {author} {\bibfnamefont {M.}~\bibnamefont {Meuwly}},\
  }\href@noop {} {\bibfield  {journal} {\bibinfo  {journal} {J. Chem. Inf.
  Model}\ }\textbf {\bibinfo {volume} {57}},\ \bibinfo {pages} {1923} (\bibinfo
  {year} {2017})}\BibitemShut {NoStop}%
\bibitem [{\citenamefont {Tennyson}\ \emph {et~al.}(2004)\citenamefont
  {Tennyson}, \citenamefont {Kostin}, \citenamefont {Barletta}, \citenamefont
  {Harris}, \citenamefont {Polyansky}, \citenamefont {Ramanlal},\ and\
  \citenamefont {Zobov}}]{ten04:85}%
  \BibitemOpen
  \bibfield  {author} {\bibinfo {author} {\bibfnamefont {J.}~\bibnamefont
  {Tennyson}}, \bibinfo {author} {\bibfnamefont {M.~A.}\ \bibnamefont
  {Kostin}}, \bibinfo {author} {\bibfnamefont {P.}~\bibnamefont {Barletta}},
  \bibinfo {author} {\bibfnamefont {G.~J.}\ \bibnamefont {Harris}}, \bibinfo
  {author} {\bibfnamefont {O.~L.}\ \bibnamefont {Polyansky}}, \bibinfo {author}
  {\bibfnamefont {J.}~\bibnamefont {Ramanlal}}, \ and\ \bibinfo {author}
  {\bibfnamefont {N.~F.}\ \bibnamefont {Zobov}},\ }\href@noop {} {\bibfield
  {journal} {\bibinfo  {journal} {Comput.\ Phys.\ Commun.}\ }\textbf {\bibinfo
  {volume} {163}},\ \bibinfo {pages} {85} (\bibinfo {year} {2004})}\BibitemShut
  {NoStop}%
\bibitem [{\citenamefont {Davidson}(1975)}]{davidson:75}%
  \BibitemOpen
  \bibfield  {author} {\bibinfo {author} {\bibfnamefont {E.~R.}\ \bibnamefont
  {Davidson}},\ }\href@noop {} {\bibfield  {journal} {\bibinfo  {journal} {J.
  Comput. Phys.}\ }\textbf {\bibinfo {volume} {17}},\ \bibinfo {pages} {87}
  (\bibinfo {year} {1975})}\BibitemShut {NoStop}%
\bibitem [{\citenamefont {Groenenboom}\ and\ \citenamefont
  {Colbert}(1993)}]{groenenboom:93b}%
  \BibitemOpen
  \bibfield  {author} {\bibinfo {author} {\bibfnamefont {G.~C.}\ \bibnamefont
  {Groenenboom}}\ and\ \bibinfo {author} {\bibfnamefont {D.~T.}\ \bibnamefont
  {Colbert}},\ }\href {\doibase 10.1063/1.465450} {\bibfield  {journal}
  {\bibinfo  {journal} {J. Chem. Phys.}\ }\textbf {\bibinfo {volume} {99}},\
  \bibinfo {pages} {9681} (\bibinfo {year} {1993})}\BibitemShut {NoStop}%
\bibitem [{\citenamefont {Koner}\ \emph {et~al.}(2016)\citenamefont {Koner},
  \citenamefont {Barrios}, \citenamefont {Gonz\'{a}lez-Lezana},\ and\
  \citenamefont {Panda}}]{kon16:034303}%
  \BibitemOpen
  \bibfield  {author} {\bibinfo {author} {\bibfnamefont {D.}~\bibnamefont
  {Koner}}, \bibinfo {author} {\bibfnamefont {L.}~\bibnamefont {Barrios}},
  \bibinfo {author} {\bibfnamefont {T.}~\bibnamefont {Gonz\'{a}lez-Lezana}}, \
  and\ \bibinfo {author} {\bibfnamefont {A.~N.}\ \bibnamefont {Panda}},\
  }\href@noop {} {\bibfield  {journal} {\bibinfo  {journal} {J.~Chem.\ Phys.}\
  }\textbf {\bibinfo {volume} {144}},\ \bibinfo {pages} {034303} (\bibinfo
  {year} {2016})}\BibitemShut {NoStop}%
\end{thebibliography}%

\end{document}


\maketitle
\begin{table}
 \caption{Grid for the diatomic potentials. All values in atomic
   units.}
 \begin{tabular}{lcccc}
 \\
 \hline
 \hline
     Method & \ \ \ \ diatom  \ \ \ \ & $N$ & \ \ \ \   $r_{\rm min}$ \ \ \ \  &  \ \ \ \ $r_{\rm max}$ \\
     \hline
    MRCI+Q & H$_2^+$ & 96& 0.8& 50.0 \\
    FCI & H$_2^+$ & 96& 0.8& 50.0 \\
    \hline
     MRCI+Q & HeH$^+$ & 102& 0.7& 50.0 \\
    FCI & HeH$^+$ & 102& 0.7& 50.0 \\
 \hline
\hline
 \end{tabular}
 \label{sitab:diat-grid}
 \end{table}

\begin{table}
\caption{Grid for the triatomic potentials. All values in atomic
  units.}
\begin{tabular}{lcc}
\hline
\hline
    & \ \ \ MRCI+Q \ \ \ & \ \ \ FCI \ \ \ \\
    \hline
    $N_{R}$ & 26 & 31 \\
    $R_{\rm min}$ & 0.3 & 1.5 \\
    $R_{\rm max}$ & 8.3 & 40.0 \\
    $N_{r}$ & 27 & 16  \\
    $r_{\rm min}$ & 0.9 & 0.9 \\
    $r_{\rm max}$ & 4.5 & 3.9 \\
    $N_{\theta}$ & 11 & 9 \\
\hline
\hline
\end{tabular}
\label{sitab:triat-grid}
\end{table}

\clearpage
\begin{table}
\caption{Coefficients for the MRCI+Q/aug-cc-pV6Z diatomic potentials.}
\begin{tabular}{lcc}
\hline
\hline
Coefficients & H$_2^+$ & HeH$^+$ \\
\hline
$c_0$   &  1.0185817913777047 &  3.2233398865576910 \\
$\alpha$&  1.6426016158795942 &  2.8140238576362075 \\
$c_1$   &-0.73688350635547695 & -3.3885062843889191 \\
$c_2$   &  5.1706704100481042 &  86.578369632144941 \\
$c_3$   & -83.346600494014567 & -2929.2064910602976 \\
$c_4$   &  874.72395010350556 &  73876.513409500127 \\
$c_5$   & -5621.3191308852929 & -1280675.3543930338 \\
$c_6$   &  23054.489047468251 &  14775779.204776313 \\
$c_7$   & -60955.753058022863 & -110729495.67500649 \\
$c_8$   &   100818.87103555651&   512524888.74876255\\
$c_9$   &  -94989.664045708429&  -1315519328.5011687\\
$c_{10}$  &   38921.331029687601&   1401896260.3220801\\
$\beta$ &  0.99809657513307815&   2.1737434370604722\\
\hline
\hline
\end{tabular}
\label{sitab:mrci-diat-coeff}
\end{table}

\clearpage
\begin{table}
\caption{Coefficients for the FCI/aug-cc-pV5Z diatomic potentials.}
\begin{tabular}{lcc}
\hline
\hline
Coefficients & H$_2^+$ & HeH$^+$ \\
\hline
$c_0$ &  1.0187329739360240 &  1.4607052781134291  \\
$\alpha$ &  1.6427288815351817 &  2.8820010945364949  \\
$c_1$ &-0.73635966940340725 & -3.4215724179324090  \\
$c_2$ &  5.1598925267102205 &  96.452411178877199  \\
$c_3$ & -83.517326447967520 & -3304.8226974189261  \\
$c_4$ &  881.03015485244248 &  89711.319427535549  \\
$c_5$ & -5696.2664864874414 & -1677314.4764108122  \\
$c_6$ &  23534.011016663684 &  21179550.008693457  \\
$c_7$ & -62775.390381188619 & -176458143.78657210  \\
$c_8$ &  104921.46142264079 &  928389438.74672687  \\
$c_9$ & -100077.26031466493 & -2792961078.7000194  \\
$c_10$ &  41598.163770554645 &  3667131828.9247351  \\
$\beta$ & 0.99821070143869606 &  2.1770149125762095  \\
\hline
\hline
\end{tabular}
\label{sitab:fci-diat-coeff}
\end{table}

\clearpage

\begin{figure}
\includegraphics[scale=0.7]{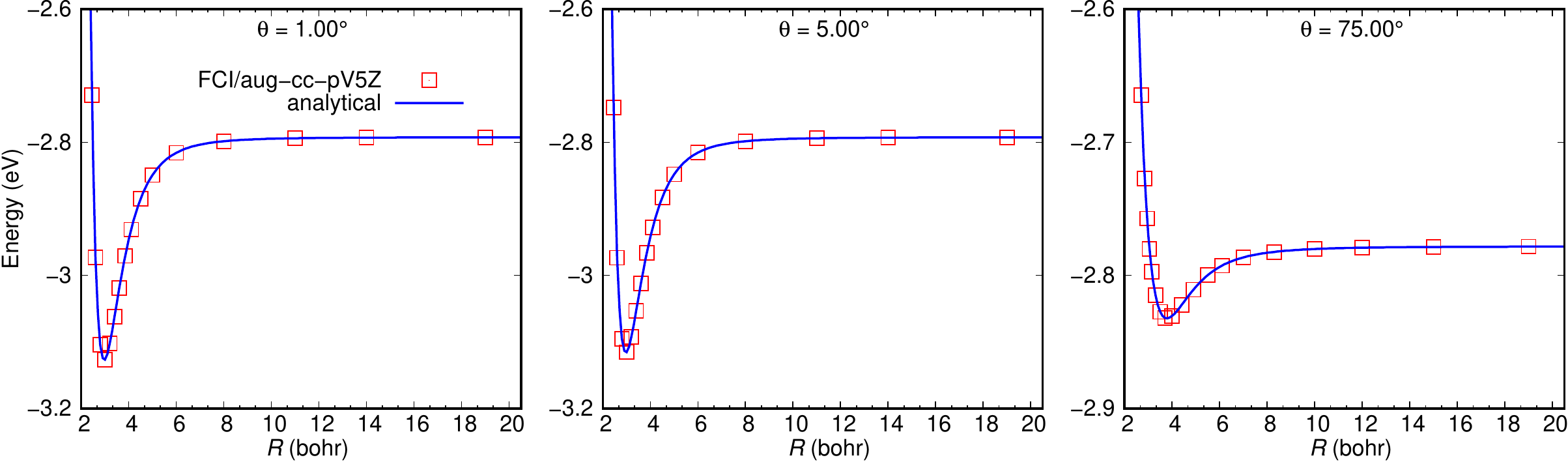}
\caption{The RKHS energies (solid lines) as a function of $R$ along
  with the FCI/aug-cc-pV5Z energies for several Jacobi angles at fixed
  $r = 2.01$ a$_0$ for $\theta=1.0, 9.0$ and $r = 1.9$ for
  $\theta=75.0$.}
\label{fig:sifig3}
\end{figure}

\begin{figure}
\includegraphics[scale=0.8]{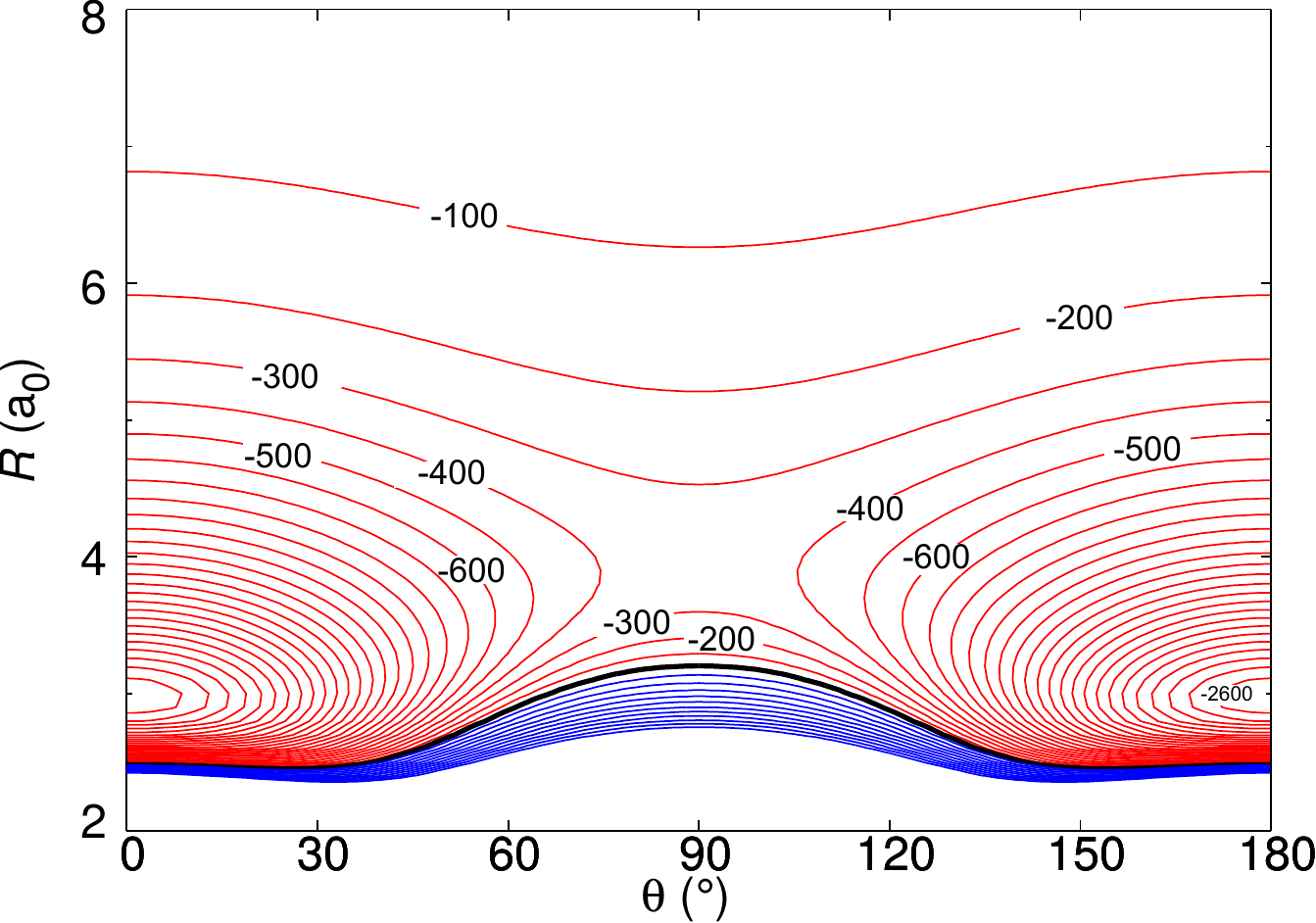}
\caption{Contour diagram for the MRCI+Q+LR PES for fixed $r = 2.0$
  a$_0$. Energies are in cm$^{-1}$ and the contour spacing is 100
  cm$^{-1}$. The energy of $r = 2.0$ a$_0$ and $R = \infty$ is the
  zero of energy. The red and blue lines correspond to negative and
  positive energies, respectively. The black line is the zero of
  energy, set to He+H$_2^+$ asymptote.}
\label{fig:sifig1}
\end{figure}

\begin{figure}
\includegraphics[scale=1.0]{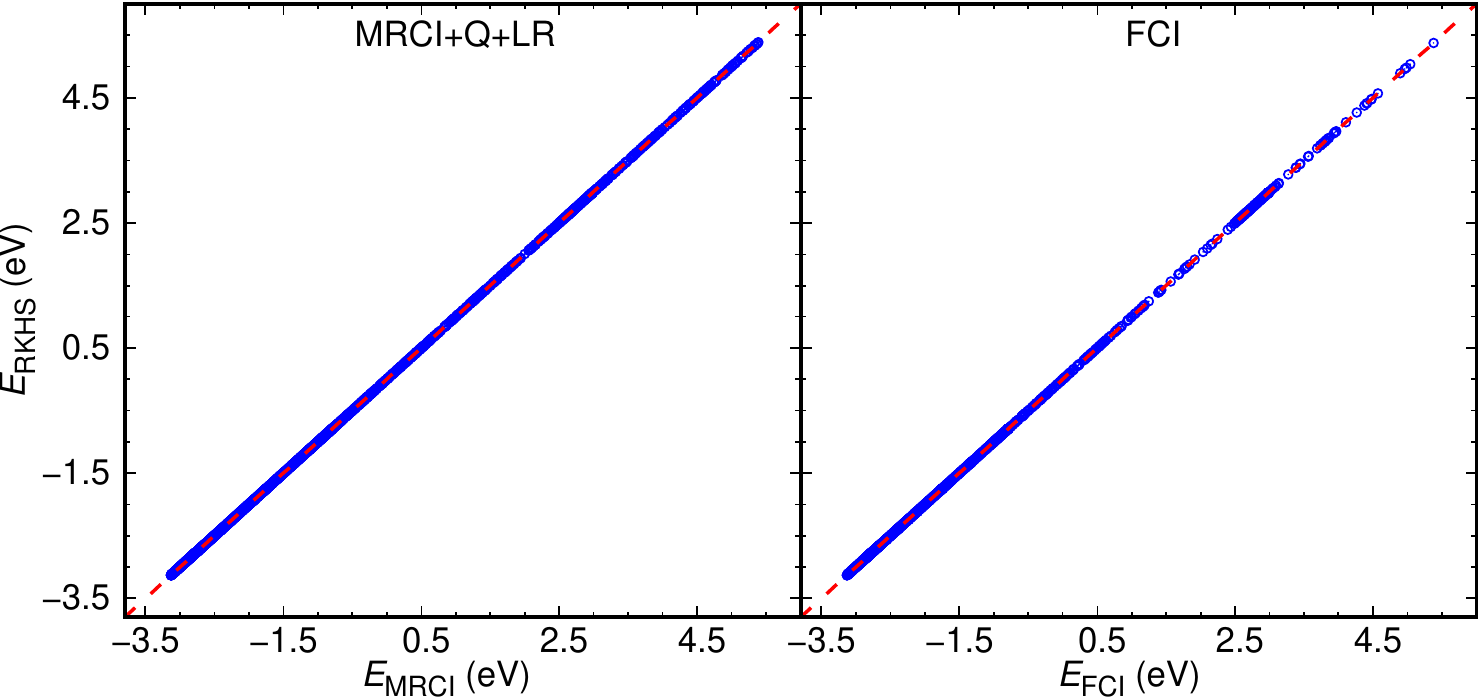}
\caption{Comparison between the reference MRCI/aug-cc-pv6Z and
  FCI/aug-cc-pv5Z {\it ab initio} on-grid energies and those obtained
  from the RKHS surfaces. The zero of energy is at $E$(H) + $E$(He),
  i.e for full dissociation.}
\label{fig:sifig2a}
\end{figure}

\begin{figure}
\includegraphics[scale=1.0]{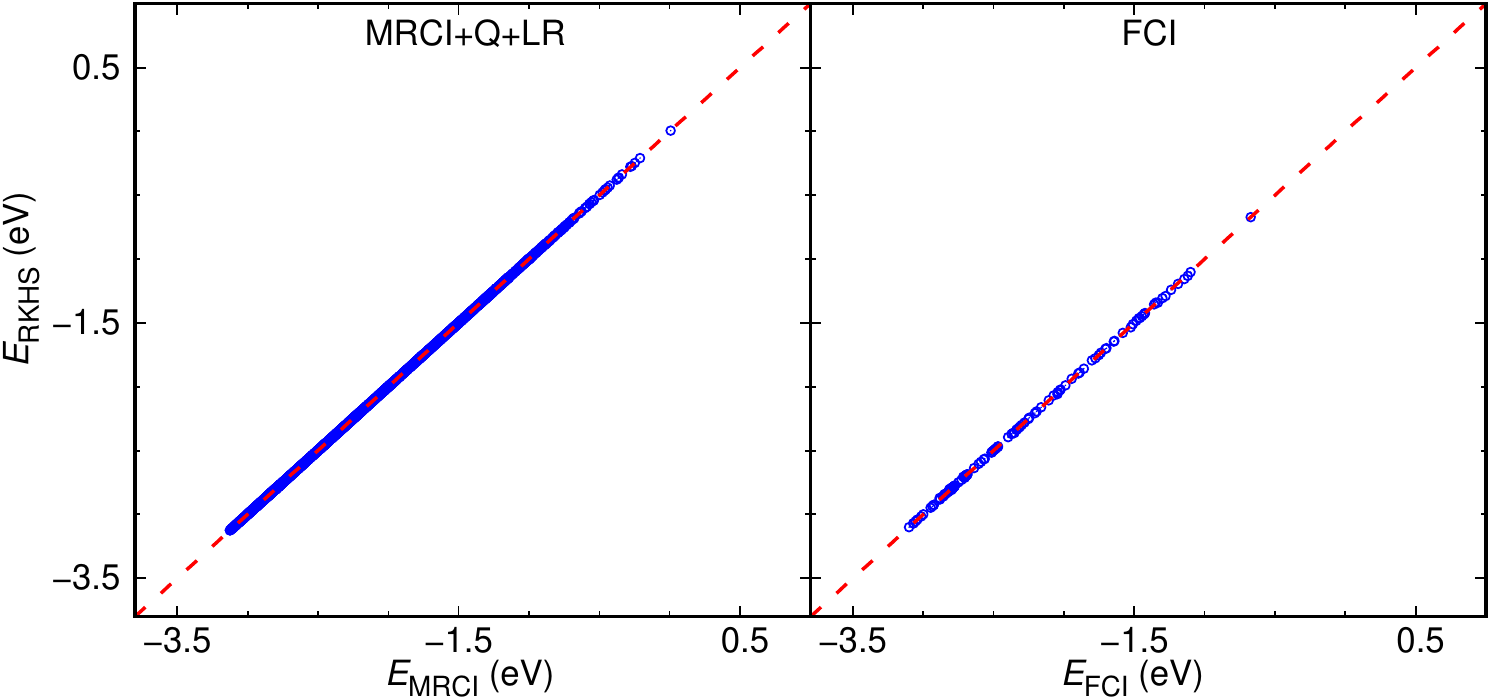}
\caption{Comparison between the reference MRCI/aug-cc-pv6Z (left) and
  FCI/aug-cc-pv5Z (right) {\it ab initio} off-grid energies and those
  obtained from the RKHS surfaces. 4912 and 146 off-grid energies for
  the MRCI/aug-cc-pv6 and FCI/aug-cc-pv5Z, respectively.  The zero of
  energy is at $E$(H) + $E$(He), i.e for full dissociation.}
\label{fig:sifig2b}
\end{figure}

\begin{figure}
\includegraphics[scale=0.6]{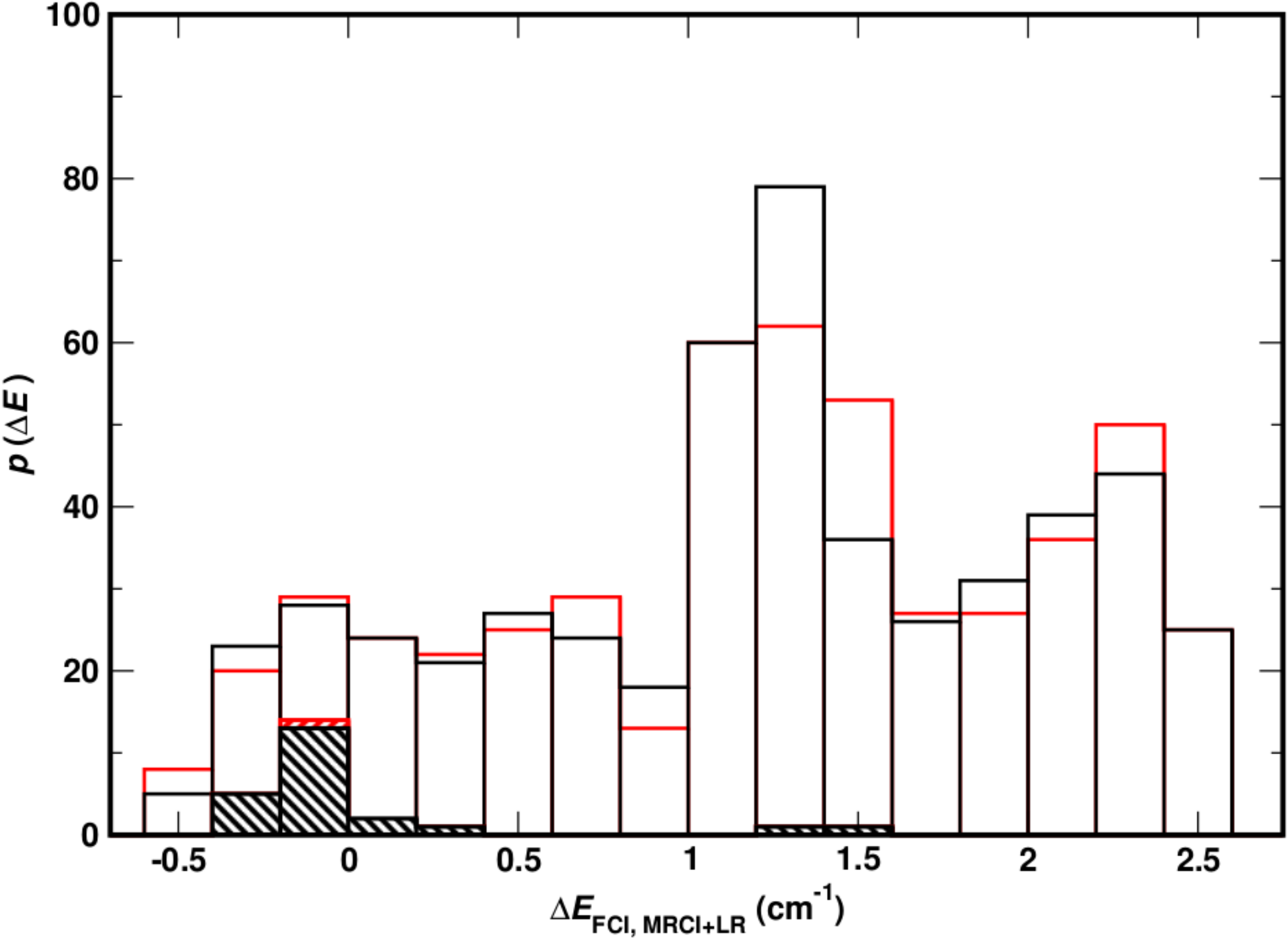}
\caption{The distribution of the energy difference $\Delta E$ on the
  FCI and MRCI+LR PESs for all bound states (i.e. $J = 0$ to 21, for
  $e-$ and $f-$parity {\it ortho} and {\it para} H$_2^+$--He). Bound
  state calculations with DVR3D (black) and CCVM (red) are reported
  separately. Open boxes for all bound states and dashed boxes for
  near-dissociative states, i.e.  within 10 cm$^{-1}$ of
  dissociation. The average difference between FCI and MRCI+LR is 1.21
  cm$^{-1}$ for both bound state methods.}
\label{fig:sifig4}
\end{figure}

\clearpage

\begin{small}

\begin{longtable}{ccccrrrr}
\caption{All bound energy levels from DVR3D and CCVM calculations for
  $e-$ and $f-$parity, {\it ortho} (o) and {\it para} (p)
  H$_2^+$--He in cm$^{-1}$, using the FCI and MRCI+Q+LR PES
  (MRCI). The zero of energy is that of H$_2^+(v=0,j=0)$.}\\
\hline
\ \ o/p \ \ &  $J$ \ \ & \ \  parity \ \  & \ \ $n$ \ \  & MRCI/DVR3D  &  \ \ FCI/DVR3D &  \ \ MRCI/CCVM &  \ \ FCI/CCVM \\
\hline
\endfirsthead
\hline
 \ \ o/p \ \ & \ \  $J$ \ \ & \ \  parity \ \  & \ \ $n$ \ \  & MRCI/DVR3D  &  \ \ FCI/DVR3D &  \ \ MRCI/CCVM &  \ \ FCI/CCVM \\
\hline
\endhead
\hline
\endfoot
\hline
\hline
\endlastfoot
o &  0  & e &  1 &  -1795.1567 & -1793.7632 & -1795.3328  & -1793.9067  \\
o &  0  & e &  2 &  -1063.7471 & -1062.6727 & -1063.3612  & -1062.2742  \\
o &  0  & e &  3 &   -660.1392 &  -657.6932 & -660.2838  &  -657.8235  \\
o &  0  & e &  4 &   -538.7132 &  -538.1281 & -538.1236  &  -537.5470  \\
o &  0  & e &  5 &   -254.8020 &  -254.3251 & -254.6102  &  -254.1331  \\
o &  0  & e &  6 &   -190.6043 &  -190.0415 & -190.4699  &  -189.9211  \\
o &  0  & e &  7 &    -48.1814 &   -48.5279 & -47.9117   &   -48.2657  \\
o &  0  & e &  8 &     19.2177 &    19.0868 & 18.5291    &    18.4236  \\
o &  0  & e &  9 &     40.3210 &    40.0360 & 39.8955    &    39.5748  \\
o &  0  & e & 10 &     55.8761 &    55.8602 &  55.6889   &    55.6620  \\ 
\hline                                                       
o &  1 &  e &  1 &  -1786.9898 & -1785.5973 & -1787.1676  & -1785.7423  \\
o &  1 &  e &  2 &  -1156.0232 & -1153.7813 & -1156.1818  & -1153.9126  \\
o &  1 &  e &  3 &  -1056.3771 & -1055.3035 & -1055.9973  & -1054.9108  \\
o &  1 &  e &  4 &   -652.4661 &  -650.0284 &  -652.6083  &  -650.1564  \\
o &  1 &  e &  5 &   -566.0520 &  -564.6730 &  -565.9535  &  -564.5672  \\
o &  1 &  e &  6 &   -531.6576 &  -531.0451 & -531.0936  &  -530.4897  \\
o &  1 &  e &  7 &   -264.3738 &  -263.2100 & -264.3214  &  -263.1671  \\
o &  1 &  e &  8 &   -249.3284 &  -248.8408 & -249.1223  &  -248.6344  \\
o &  1 &  e &  9 &   -186.9282 &  -186.1513 & -186.8255  &  -186.0576  \\
o &  1 &  e & 10 &   -175.7091 &  -174.2152 & -175.7432  &  -174.2595  \\
o &  1 &  e & 11 &    -69.2871 &   -68.9712 & -69.1863  &   -68.8806  \\
o &  1 &  e & 12 &    -44.7877 &   -45.1411 & -44.5115  &   -44.8730  \\
o &  1 &  e & 13 &    -27.7742 &   -27.0299 & -27.8145  &   -27.0795  \\
o &  1 &  e & 14 &     23.0413 &    22.8241 &  22.5508  &    22.3559  \\
o &  1 &  e & 15 &     30.9127 &    30.5164 &  30.8918  &    30.4832  \\
o &  1 &  e & 16 &     43.6677 &    43.4379 &  43.1988  &    42.9343  \\
o &  1 &  e & 17 &     54.5821 &    54.5753 &  54.5141  &    54.4935  \\
o &  1 &  e & 18 &     57.2725 &    57.2317 &  56.9590  &    56.9161  \\
o &  1 &  f &  1 &  -1155.9068 & -1153.4648 & -1155.8671  & -1153.5957  \\
o &  1 &  f &  2 &   -565.3280 &  -563.7240 &  -565.0456  &  -563.6327  \\
o &  1 &  f &  3 &   -264.2372 &  -262.8397 & -263.9824  &  -262.7923  \\
o &  1 &  f &  4 &   -177.7497 &  -175.8616 & -177.6001  &  -175.9153  \\
o &  1 &  f &  5 &    -69.2720 &   -68.7581 & -68.9665  &   -68.6609  \\
o &  1 &  f &  6 &    -28.4535 &   -27.5094 & -28.2910  &   -27.5540  \\
o &  1 &  f &  7 &     30.1113 &    29.9118 &  30.3342  &    29.9213  \\
o &  1 &  f &  8 &     54.4640 &    54.6476 &  54.6289  &    54.6006  \\
\hline                                                       
o &  2 &  e &  1 &  -1770.6650 & -1769.2743 & -1770.8461  & -1769.4226  \\
o &  2 &  e &  2 &  -1140.2714 & -1138.0381 & -1140.4328  & -1138.1719  \\
o &  2 &  e &  3 &  -1041.6441 & -1040.5717 & -1041.2764  & -1040.1910  \\
o &  2 &  e &  4 &   -637.1448 &  -634.7245 &  -637.2830  &  -634.8481  \\
o &  2 &  e &  5 &   -557.3420 &  -555.7147 &  -557.2793  &  -555.6344  \\
o &  2 &  e &  6 &   -541.3563 &  -538.7918 &  -541.4221  &  -538.8453  \\
o &  2 &  e &  7 &   -517.4518 &  -516.7724 &  -516.9418  &  -516.2723  \\
o &  2 &  e &  8 &   -253.2841 &  -252.2489 &  -253.2366  &  -252.2108  \\
o &  2 &  e &  9 &   -238.2869 &  -237.7936 &  -238.0550  &  -237.5618  \\
o &  2 &  e & 10 &   -177.7376 &  -176.8303 &  -177.6672  &  -176.7689  \\
o &  2 &  e & 11 &   -160.9265 &  -159.4781 &  -160.9730  &  -159.5339  \\
o &  2 &  e & 12 &   -129.0765 &  -127.7549 &  -129.0885  &  -127.7639  \\
o &  2 &  e & 13 & -60.8852 &   -60.6194    &   -60.7985  &   -60.5431  \\
o &  2 &  e & 14 & -38.1052 &   -38.4707    &   -37.8218  &   -38.1967  \\
o &  2 &  e & 15 & -16.2651 &   -15.5362    &   -16.3163  &   -15.5962  \\
o &  2 &  e & 16 &  29.2257 &    28.9067    &    29.0301  &    28.7185  \\
o &  2 &  e & 17 &  36.8861 &    36.4658    &    36.8443  &    36.4166  \\
o &  2 &  e & 18 &  49.7546 &    49.6466    &    49.3446  &    49.2128  \\
o &  2 &  e & 19 &  56.9232 &    56.9159    &    56.6388  &    56.6233  \\
o &  2 &  e & 20 &  59.1856 &    59.1620    &    58.2724  &    58.2708  \\
o &  2 &  f &  1 & 139.3222 & -1137.0820    & -1139.4823  & -1137.2146  \\
o &  2 &  f &  2 & 555.4666 &  -553.6852    &   -555.4432 &  -553.6438  \\
o &  2 &  f &  3 & 540.3615 &  -537.8587    &   -540.4319 &  -537.9191  \\
o &  2 &  f &  4 & 252.0531 &  -250.9242    &   -251.9933 &  -250.8740  \\
o &  2 &  f &  5 & 164.7003 &  -162.9638    &   -164.7550 &  -163.0232  \\
o &  2 &  f &  6 & 129.1048 &  -127.7836    &   -129.1167 &  -127.7924  \\
o &  2 &  f &  7 & -60.2646 &   -60.0011    &    -60.1577 &   -59.9049  \\
o &  2 &  f &  8 & -17.6055 &   -16.8608    &    -17.6396 &   -16.9030  \\
o &  2 &  f &  9 &  34.9327 &    34.5617    &     34.9539 &    34.5675  \\
o &  2 &  f & 10 &  56.4390 &    56.4365    &     56.3285 &    56.3143  \\
\hline                                                     
o &  3 &  e &  1 & 746.1992 & -1744.8124    & -1746.3862   & -1744.9654  \\
o &  3 &  e &  2 & 116.6904 & -1114.4702    & -1116.8561   & -1114.6083  \\
o &  3 &  e &  3 & 019.5579 & -1018.4905    & -1019.2123   & -1018.1278  \\
o &  3 &  e &  4 & 614.2254 &  -611.8344    &  -614.3605   &  -611.9526  \\
o &  3 &  e &  5 & 540.9512 &  -539.3824    &  -540.8797   &  -539.2946  \\
o &  3 &  e &  6 & 516.6202 &  -514.1562    &  -516.6592   &  -514.1774  \\
o &  3 &  e &  7 & 495.8888 &  -495.0796    &  -495.4694   &  -494.6700  \\
o &  3 &  e &  8 & 237.3570 &  -236.4672    &  -237.3158   &  -236.4336  \\
o &  3 &  e &  9 & 221.7092 &  -221.2521    &  -221.4532   &  -220.9932  \\
o &  3 &  e & 10 & 162.4168 &  -161.4340    &  -162.3956   &  -161.4212  \\
o &  3 &  e & 11 & 139.8134 &  -138.3285    &  -139.8837   &  -138.4064  \\
o &  3 &  e & 12 & 111.6749 &  -110.3976    &  -111.6955   &  -110.4156  \\
o &  3 &  e & 13 & -48.6143 &   -48.4085    &   -48.5513   &   -48.3555  \\
o &  3 &  e & 14 & -29.6395 &   -28.6943    &   -29.6833   &   -28.4228  \\
o &  3 &  e & 15 & -28.3078 &   -27.3409    &   -28.0281   &   -27.3808  \\
o &  3 &  e & 16 &   0.4480 &     1.0877    &     0.3795   &     1.0112  \\
o &  3 &  e & 17 &  36.9032 &    36.5421    &    36.8878   &    36.5234  \\
o &  3 &  e & 18 &  44.9410 &    44.5130    &    44.9380   &    44.4973  \\
o &  3 &  e & 19 &  55.5880 &    55.5770    &    55.4434   &    55.4216  \\
o &  3 &  f &  1 & 114.7727 & -1112.5383    & -1114.9357   & -1112.6735  \\
o &  3 &  f &  2 & 537.4321 &  -535.6310    &  -537.4179   &  -535.6006  \\
o &  3 &  f &  3 & 514.0733 &  -511.6189    &  -514.1462   &  -511.6795  \\
o &  3 &  f &  4 & 234.4370 &  -233.4051    &  -234.3742   &  -233.3524  \\
o &  3 &  f &  5 & 145.1540 &  -143.3639    &  -145.2173   &  -143.4310  \\
o &  3 &  f &  6 & 111.8198 &  -110.5442    &  -111.8399   &  -110.5617  \\
o &  3 &  f &  7 & -47.4665 &   -47.2688    &   -47.3632   &   -47.1765  \\
o &  3 &  f &  8 & -29.6378 &   -27.3421    &   -29.6818   &   -27.3825  \\
o &  3 &  f &  9 &  -1.7765 &    -1.0725    &    -1.8056   &    -1.1087  \\
o &  3 &  f & 10 &  41.7827 &    41.4734    &   41.7976    &    41.4729  \\
\hline                                          
o &  4 &  e &  1 & -1713.6215 & -1712.2384 & -1713.8150  & -1712.3977  \\  
o &  4 &  e &  2 & -1085.3377 & -1083.1359 & -1085.5099  & -1083.2803  \\  
o &  4 &  e &  3 &  -990.1450 &  -989.0786 &  -989.8239  &  -988.7402  \\  
o &  4 &  e &  4 &  -583.7937 &  -581.4420 &  -583.9254  &  -581.5562  \\  
o &  4 &  e &  5 &  -517.8296 &  -516.3664 &  -517.7529  &  -516.2752  \\  
o &  4 &  e &  6 &  -485.6063 &  -483.3000 &  -485.6049  &  -483.2746  \\  
o &  4 &  e &  7 &  -466.6638 &  -465.6226 &  -466.3657  &  -465.3387  \\  
o &  4 &  e &  8 &  -216.8848 &  -216.1171 &  -216.8498  &  -216.0881  \\  
o &  4 &  e &  9 &  -200.0454 &  -199.6724 &  -199.7703  &  -199.3986  \\  
o &  4 &  e & 10 &  -140.8270 &  -139.8171 &  -140.8779  &  -139.8747  \\  
o &  4 &  e & 11 &  -112.1289 &  -110.5610 &  -112.2321  &  -110.6703  \\  
o &  4 &  e & 12 &   -88.6270 &   -87.4017 &   -88.6600  &   -87.4326  \\  
o &  4 &  e & 13 &   -32.6719 &   -32.5266 &   -32.6432  &   -32.5074  \\  
o &  4 &  e & 14 &   -15.6217 &   -16.0210 &   -15.3586  &   -15.7680  \\  
o &  4 &  e & 15 &    -0.4856 &     1.7209 &    -0.5304  &     1.6795  \\  
o &  4 &  e & 16 &    20.7175 &    21.0952 &    20.6360  &    21.0072  \\  
o &  4 &  e & 17 &    45.9709 &    45.6105 &    46.0219  &    45.6523  \\  
o &  4 &  e & 18 &    54.3754 &    54.0835 &    54.3747  &    54.0665  \\  
o &  4 &  f &  1 & -1082.0973 & -1079.8707 & -1082.2640  & -1080.0094  \\            
o &  4 &  f &  2 &  -512.0117 &  -510.2382 &  -512.0047  &  -510.2162  \\            
o &  4 &  f &  3 &  -480.5780 &  -478.1346 &  -480.6601  &  -478.2027  \\            
o &  4 &  f &  4 &  -211.5633 &  -210.6444 &  -211.4995  &  -210.5912  \\            
o &  4 &  f &  5 &  -118.7145 &  -116.8834 &  -118.7874  &  -116.9592  \\ 
o &  4 &  f &  6 &   -89.0727 &   -87.8501 &   -89.1049  &   -87.8802  \\ 
o &  4 &  f &  7 &   -31.0142 &   -30.8839 &   -30.9189  &   -30.7998  \\ 
o &  4 &  f &  8 &    -0.4857 &     1.7205 &    -0.5302  &     1.6793  \\ 
o &  4 &  f &  9 &    18.4858 &    19.0399 &    18.4686  &    19.0165  \\ 
o &  4 &  f & 10 &    50.7116 &    50.5470 &   50.7177   &    50.5378  \\   
\hline                                                            
o &  5 &  e &  1 & -1672.9671 & -1671.5886 & -1673.1685  & -1671.7557  \\   
o &  5 &  e &  2 & -1046.2925 & -1044.1152 & -1046.4749  & -1044.2695  \\   
o &  5 &  e &  3 &  -953.4240 &  -952.3572 &  -953.1348  &  -952.0505  \\   
o &  5 &  e &  4 &  -545.9660 &  -543.6687 &  -546.1009  &  -543.7852  \\   
o &  5 &  e &  5 &  -488.3004 &  -486.9574 &  -488.2432  &  -486.8887  \\   
o &  5 &  e &  6 &  -448.4808 &  -446.4055 &  -448.4368  &  -446.3358  \\   
o &  5 &  e &  7 &  -429.4606 &  -428.1031 &  -429.3026  &  -427.9588  \\   
o &  5 &  e &  8 &  -191.8566 &  -191.1840 &  -191.8381  &  -191.1693  \\   
o &  5 &  e &  9 &  -173.9936 &  -173.7302 &  -173.7344  &  -173.4777  \\   
o &  5 &  e & 10 &  -113.2100 &  -112.2307 &  -113.3529  &  -112.3778  \\   
o &  5 &  e & 11 &   -77.5370 &   -75.8753 &   -77.6898  &   -76.0327  \\   
o &  5 &  e & 12 &   -60.0243 &   -58.8556 &   -60.0797  &   -58.9100  \\   
o &  5 &  e & 13 &   -13.1595 &   -13.0714 &   -13.1826  &   -13.1026  \\   
o &  5 &  e & 14 &    -0.2589 &    -0.6713 &    -0.0448  &    -0.4701  \\   
o &  5 &  e & 15 &    35.3405 &    37.4275 &    35.2899  &    37.3794  \\   
o &  5 &  e & 16 &    40.7105 &    40.6264 &    40.6545  &    40.5654  \\   
o &  5 &  e & 17 &    55.5065 &    55.2199 &    55.5180  &    55.2194  \\   
o &  5 &  f &  1 & -1041.3455 & -1039.1287 & -1041.5171  & -1039.2719  \\       
o &  5 &  f &  2 &  -479.4488 &  -477.7225 &  -479.4520  &  -477.7116  \\              
o &  5 &  f &  3 &  -439.7929 &  -437.3517 &  -439.8871  &  -437.4305  \\              
o &  5 &  f &  4 &  -183.8249 &  -183.0232 &  -183.7660  &  -182.9753  \\              
o &  5 &  f &  5 &   -85.3406 &   -83.5124 &   -85.4221  &   -83.5960  \\              
o &  5 &  f &  6 &   -61.0697 &   -59.8972 &   -61.1301  &   -59.9564  \\              
o &  5 &  f &  7 &   -11.1566 &   -11.0850 &   -11.0753  &   -11.0152  \\              
o &  5 &  f &  8 &    35.3263 &    37.3994 &    35.2794  &    37.3553  \\   
o &  5 &  f &  9 &    40.6802 &    40.8595 &    40.6822  &    40.8541  \\     
\hline                                                             
o &  6 &  e &  1 & -1624.2818 & -1622.9089 & -1624.4923  & -1623.0847  \\     
o &  6 &  e &  2 &  -999.6579 &  -997.5123 &  -999.8573  &  -997.6835  \\     
o &  6 &  e &  3 &  -909.4215 &  -908.3524 &  -909.1702  &  -908.0831  \\     
o &  6 &  e &  4 &  -500.9106 &  -498.6890 &  -501.0626  &  -498.8211  \\     
o &  6 &  e &  5 &  -452.6194 &  -451.3956 &  -452.6219  &  -451.3897  \\     
o &  6 &  e &  6 &  -405.2982 &  -403.4527 &  -405.2359  &  -403.3702  \\     
o &  6 &  e &  7 &  -384.1933 &  -382.5218 &  -384.1644  &  -382.4986  \\     
o &  6 &  e &  8 &  -162.2267 &  -161.6416 &  -162.2261  &  -161.6416  \\     
o &  6 &  e &  9 &  -144.0906 &  -143.9203 &  -143.9110  &  -143.7529  \\     
o &  6 &  e & 10 &   -80.2280 &   -79.3342 &   -80.4695  &   -79.5773  \\     
o &  6 &  e & 11 &   -35.9887 &   -34.2869 &   -36.2191  &   -34.5191  \\     
o &  6 &  e & 12 &   -26.0048 &   -24.8945 &   -26.0839  &   -24.9747  \\     
o &  6 &  e & 13 &     9.8977 &     9.9345 &     9.8250  &     9.8592  \\     
o &  6 &  e & 14 &    17.6004 &    17.1964 &    17.7097  &    17.2862  \\     
o &  6 &  e & 15 &    55.9592 &    55.6103 &    55.9853  &    55.6260  \\     
o &  6 &  f &  1 &  -992.5800 &  -990.3749 & -992.7575  &  -990.5237  \\               
o &  6 &  f &  2 &  -439.9244 &  -438.2572 & -439.9441  &  -438.2632  \\               
o &  6 &  f &  3 &  -391.7431 &  -389.3036 & -391.8532  &  -389.3973  \\               
o &  6 &  f &  4 &  -151.5839 &  -150.8949 & -151.5372  &  -150.8593  \\               
o &  6 &  f &  5 &   -45.4746 &   -43.7665 &  -45.5574  &   -43.8511  \\               
o &  6 &  f &  6 &   -27.9696 &   -26.8052 &  -28.0677  &   -26.9022  \\               
o &  6 &  f &  7 &    12.0288 &    12.0658 &   12.0901  &    12.1160  \\               
\hline                                                              
o &  7 &  e &  1 & -1567.6204 & -1566.2541 & -1567.8414  & -1566.4400  \\    
o &  7 &  e &  2 &  -945.5651 &  -943.4588 &  -945.7937  &  -943.6603  \\    
o &  7 &  e &  3 &  -858.1611 &  -857.0861 &  -857.9544  &  -856.8607  \\    
o &  7 &  e &  4 &  -448.8730 &  -446.7622 &  -449.0682  &  -446.9342  \\    
o &  7 &  e &  5 &  -410.9804 &  -409.8523 &  -411.1120  &  -409.9840  \\    
o &  7 &  e &  6 &  -355.8919 &  -354.2128 &  -355.8510  &  -354.1584  \\    
o &  7 &  e &  7 &  -331.0639 &  -329.1483 &  -331.1412  &  -329.2230  \\    
o &  7 &  e &  8 &  -128.1155 &  -127.6363 &  -128.1281  &  -127.6448  \\  
o &  7 &  e &  9 &  -110.4851 &  -110.3598 &  -110.4777  &  -110.3725  \\  
o &  7 &  e & 10 &   -42.8309 &   -42.0637 &   -43.1758  &   -42.4089  \\  
o &  7 &  e & 11 &    10.8994 &    12.1774 &    10.6460  &    11.9612  \\  
o &  7 &  e & 12 &    13.3644 &    14.5894 &    13.1879  &    14.3806  \\  
o &  7 &  e & 13 &    35.9485 &    35.7146 &    36.0581  &    35.8310  \\  
o &  7 &  e & 14 &    39.2843 &    39.3321 &    38.9474  &    38.9564  \\  
o &  7 &  f &  1 &  -935.8766 &  -933.6843 & -936.0620  &  -933.8415  \\            
o &  7 &  f &  2 &  -393.6232 &  -392.0217 & -393.6683  &  -392.0556  \\            
o &  7 &  f &  3 &  -336.5044 &  -334.0699 & -336.6360  &  -334.1849  \\            
o &  7 &  f &  4 &  -115.1570 &  -114.5708 & -115.1308  &  -114.5572  \\            
o &  7 &  f &  5 &    -1.5438 &    -0.3402 &   -1.6030  &    -0.4049  \\            
o &  7 &  f &  6 &    11.1023 &    12.4827 &   10.9391  &    12.3239  \\            
o &  7 &  f &  7 &    38.9194 &    39.0065 &   38.9521  &    39.0264  \\              
\hline                                                                    
o &  8 &  e &  1 & -1503.0477 & -1501.6888 & -1503.2807  &  -1501.8862 \\            
o &  8 &  e &  2 &  -884.1731 &  -882.1193 &  -884.4559  &   -882.3741 \\   
o &  8 &  e &  3 &  -799.6614 &  -798.5745 &  -799.5066  &   -798.4006 \\   
o &  8 &  e &  4 &  -390.2388 &  -388.3172 &  -390.5444  &   -388.5872 \\   
o &  8 &  e &  5 &  -363.4263 &  -362.3346 &  -363.8231  &   -362.7437 \\   
o &  8 &  e &  6 &  -300.0590 &  -298.4784 &  -300.0699  &   -298.4801 \\   
o &  8 &  e &  7 &  -270.3484 &  -268.2677 &  -270.5540  &   -268.4646 \\   
o &  8 &  e &  8 &   -90.0081 &   -89.6800 &   -90.0127  &    -89.6747 \\   
o &  8 &  e &  9 &   -72.9295 &   -72.7911 &   -73.2276  &    -73.1136 \\   
o &  8 &  e & 10 &    -2.2095 &    -1.6176 &    -2.6983  &     -2.1025 \\ 
o &  8 &  e & 11 &    51.1076 &    51.1344 &    51.0256  &     51.0550 \\ 
o &  8 &  e & 12 &    56.6101 &    56.9385 &    56.3904  &     56.7232 \\ 
o &  8 &  f &  1 &  -871.3245 &  -869.1492 & -871.5213  &   -869.3164 \\           
o &  8 &  f &  2 &  -340.7447 &  -339.2200 & -340.8341  &   -339.2967 \\           
o &  8 &  f &  3 &  -274.1857 &  -271.7633 & -274.3558  &   -271.9163 \\           
o &  8 &  f &  4 &   -74.8522 &   -74.3656 &  -74.8583  &    -74.3827 \\           
o &  8 &  f &  5 &    39.2080 &    39.6492 &   39.1558  &     39.5857 \\             
\hline                                                                    
o &  9 &  e &  1 & -1430.6382 & -1429.2875 & -1430.8852  &  -1429.4986 \\           
o &  9 &  e &  2 &  -815.6793 &  -813.6903 &  -816.0584  &   -814.0418 \\           
o &  9 &  e &  3 &  -733.9299 &  -732.8231 &  -733.8371  &   -732.7108 \\           
o &  9 &  e &  4 &  -326.2806 &  -324.7805 &  -326.5207  &   -324.9581 \\           
o &  9 &  e &  5 &  -309.3565 &  -308.0815 &  -310.3422  &   -309.1092 \\           
o &  9 &  e &  6 &  -237.6805 &  -236.1484 &  -237.7629  &   -236.2233 \\  
o &  9 &  e &  7 &  -202.3919 &  -200.2098 &  -202.7129  &   -200.5177 \\  
o &  9 &  e &  8 &   -48.9250 &   -48.7919 &   -48.8680  &    -48.7272 \\  
o &  9 &  e &  9 &   -30.8462 &   -30.6475 &   -31.6123  &    -31.4351 \\  
o &  9 &  e & 10 &    39.3835 &    39.6961 &    38.8858  &     39.2048 \\  
o &  9 &  f &  1 &  -799.0312 &  -796.8741 & -799.2422  &   -797.0550 \\                 
o &  9 &  f &  2 &  -281.5350 &  -280.0931 & -281.6835  &   -280.2287 \\                
o &  9 &  f &  3 &  -204.9543 &  -202.5534 & -205.1775  &   -202.7589 \\                
o &  9 &  f &  4 &   -31.0556 &   -30.6663 &  -31.1030  &    -30.7242 \\                
\hline                                                                    
o & 10 &  e &  1 & -1350.4761 & -1349.1343 & -1350.7409  &  -1349.3629 \\                
o & 10 &  e &  2 &  -740.3013 &  -738.3926 &  -740.8643  &   -738.9295 \\                 
o & 10 &  e &  3 &  -660.9611 &  -659.8237 &  -660.9438  &   -659.7868 \\                 
o & 10 &  e &  4 &  -260.8401 &  -260.0968 &  -260.8683  &   -260.1190 \\                 
o & 10 &  e &  5 &  -245.4741 &  -243.6986 &  -247.5271  &   -245.7409 \\                 
o & 10 &  e &  6 &  -168.7289 &  -167.2100 &  -168.9012  &   -167.3766 \\                 
o & 10 &  e &  7 &  -127.4583 &  -125.2220 &  -127.9104  &   -125.6595 \\                 
o & 10 &  e &  8 &    -6.0023 &    -6.0529 &    -5.9376  &     -5.9911 \\            
o & 10 &  e &  9 &    16.1354 &    16.3716 &    14.8284  &     15.0546 \\   
o & 10 &  f &  1 &  -719.1130 &  -716.9760 & -719.3481  &   -717.1809 \\                 
o & 10 &  f &  2 &  -216.2559 &  -214.9037 & -216.5030  &   -215.1385 \\                 
o & 10 &  f &  3 &  -129.0195 &  -126.6502 & -129.3216  &   -126.9357 \\                 
o & 10 &  f &  4 &    15.6765 &    15.9619 &   15.5749  &     15.8500 \\                
\hline                                                                    
o & 11 &  e &  1 & -1262.6767 & -1261.3449 & -1262.9448  &  -1261.5763 \\                 
o & 11 &  e &  2 &  -658.8037 &  -656.9979 &  -659.1889  &   -657.3550 \\                 
o & 11 &  e &  3 &  -580.7891 &  -579.6097 &  -580.8132  &   -579.6132 \\                 
o & 11 &  e &  4 &  -196.5935 &  -196.1791 &  -196.5529  &   -196.1374 \\                 
o & 11 &  e &  5 &  -172.1815 &  -170.3586 &  -173.2863  &   -171.4489 \\                 
o & 11 &  e &  6 &   -93.3707 &   -91.8494 &   -93.5607  &    -92.0304 \\                 
o & 11 &  e &  7 &   -46.1749 &   -43.9271 &   -46.4655  &    -44.2000 \\                 
o & 11 &  e &  8 &    37.7768 &    37.5902 &    37.9437  &     37.7480 \\                 
o & 11 &  f &  1 &  -631.7572 &  -629.6438 & -631.9812  &   -629.8367 \\                         
o & 11 &  f &  2 &  -145.4933 &  -144.2397 & -145.6397  &   -144.3744 \\                  
o & 11 &  f &  3 &   -46.8659 &   -44.5473 &  -47.0835  &    -44.7459 \\                
\hline                                                                      
o & 12 &  e &  1 & -1167.3192 & -1165.9982 & -1167.6061  &  -1166.2479 \\                   
o & 12 &  e &  2 &  -570.8360 &  -569.1503 &  -571.3966  &   -569.6836 \\                  
o & 12 &  e &  3 &  -493.3133 &  -492.0779 &  -493.4302  &   -492.1746 \\                 
o & 12 &  e &  4 &  -129.5925 &  -129.3066 &  -129.8373  &   -129.5459 \\                 
o & 12 &  e &  5 &   -90.7462 &   -89.1006 &   -92.4048  &    -90.7492 \\                 
o & 12 &  e &  6 &   -11.7334 &   -10.2055 &   -12.0204  &    -10.4858 \\                   
o & 12 &  e &  7 &    41.6331 &    43.8562 &    41.2143  &     43.4554 \\                   
o & 12 &  f &  1 &  -537.0590 &  -534.9714 & -537.3052  &   -535.1862 \\                        
o & 12 &  f &  2 &   -69.3031 &   -68.1596 &  -69.5283  &    -68.3736 \\                                        
o & 12 &  f &  3 &    41.4141 &    43.6637 &   41.1253  &     43.3933 \\                                             
\hline                                                                                            
o & 13 &  e &  1 & -1064.5293 & -1063.2198 & -1064.8471  &  -1063.5000 \\                               
o & 13 &  e &  2 &  -477.0168 &  -475.4694 &  -477.8862  &   -476.3124 \\                               
o & 13 &  e &  3 &  -398.5395 &  -397.2364 &  -398.8007  &   -397.4797 \\                               
o & 13 &  e &  4 &   -59.1220 &   -58.9456 &   -59.7471  &    -59.5620 \\                              
o & 13 &  e &  5 &    -4.6681 &    -3.2424 &    -7.1159  &     -5.6813 \\                              
o & 13 &  f &  1 &  -435.2213 &  -433.1624 & -435.5084  &   -433.4183 \\                                         
o & 13 &  f &  2 &    11.6097 &    12.6276 &   11.2553  &     12.2837 \\  
\hline
o & 14 &  e &  1 & -954.4943 & -953.1976 & -954.8042  &  -953.4693\\
o & 14 &  e &  2 & -378.5088 & -377.1159 & -379.0680  &  -377.6469\\
o & 14 &  e &  3 & -296.7142 & -295.3452 & -296.9874  &  -295.5975\\
o & 14 &  e &  4 &   13.6474 &   13.7187 &   13.3700  &    13.4468\\
o & 14 &  f &  1 & -326.5530 & -324.5282 & -326.8090  &   -324.7518 \\ 
\hline
o & 15 &  e &  1 & -837.2827 & -835.9977 & -837.6304  &  -836.3087\\
o & 15 &  e &  2 & -274.4768 & -273.2405 & -275.3434  &  -274.0819\\
o & 15 &  e &  3 & -187.6883 & -186.2518 & -188.1490  &  -186.6951\\
o & 15 &  f &  1 & -211.1674 & -209.1802 & -211.4621 &   -209.4427 \\ 
\hline
o & 16 &  e &  1 & -713.0808 & -711.8117 & -713.4964 &   -712.1891 \\
o & 16 &  e &  2 & -165.7499 & -164.6763 & -167.1046 &   -166.0040 \\
o & 16 &  e &  3 &  -71.7692 &  -70.2764 &  -72.5696 &    -71.0656 \\
o & 16 &  f &  1 &  -89.4000 & -87.4563 & -89.7708&      -87.7954 \\ 
\hline
o & 17 &  e &  1 & -582.0521 & -580.7982 & -582.5945 & -581.3031 \\
o & 17 &  e &  2 &  -52.6909 &  -51.7802 &  -54.7654 &  -53.8244 \\
o & 17 &  e &  3 &   50.8555 &   52.3881 &   49.3175 &   50.8505 \\
o & 17 &  f &  1 &   38.4136 &   40.3062 &   37.8960 &   39.8191 \\ 
\hline
o & 18 &  e &  1 & -444.3786 & -443.1407 & -445.1424 &  -443.8686 \\
\hline
o & 19 &  e &  1 & -300.2620 & -299.0414 & -301.3889 &  -300.1350 \\
\hline
o & 20 &  e &  1 & -149.7211 & -148.5150 & -151.6224 &  -150.3913 \\
\hline
o & 21 &  e &  1 & 6.7592 & 7.9412 & 3.8163 &  5.0204 \\
\hline
\hline
  p &  0 &  e &  1 & -1795.1575    & -1793.7639  &  -1795.3352  &   -1793.9091  \\                                         
  p &  0 &  e &  2 & -1063.7827    & -1062.7083  &  -1063.3984  &   -1062.3114  \\                                        
  p &  0 &  e &  3 &  -662.1132    &  -659.6723  &   -662.2749  &    -659.8206  \\                                        
  p &  0 &  e &  4 &  -539.4297    &  -538.8389  &   -538.8525  &    -538.2706  \\                                        
  p &  0 &  e &  5 &  -301.4467    &  -300.6517  &   -301.4014  &    -300.6175  \\                                        
  p &  0 &  e &  6 &  -214.4946    &  -214.3129  &   -214.1766  &    -214.0097  \\                                        
  p &  0 &  e &  7 &  -109.9271    &  -109.8684  &   -109.8808  &    -109.8289  \\                                        
  p &  0 &  e &  8 &   -60.6972    &   -60.7249  &    -60.7455  &     -60.7929  \\                                        
  p &  0 &  e &  9 &   -15.7192    &   -16.0670  &    -15.6409  &     -16.0053  \\                                        
  p &  0 &  e & 10 &    -1.1368    &    -1.1504  &     -1.5090  &      -1.5265  \\                                        
  \hline 
  p &  1 &  e &  1 & -1786.9905    & -1785.5980  &  -1787.1699  &   -1785.7447  \\                                        
  p &  1 &  e &  2 & -1155.9897    & -1153.7480  &  -1156.1568  &   -1153.8879  \\                                        
  p &  1 &  e &  3 & -1056.4101    & -1055.3364  &  -1056.0319  &   -1054.9454  \\                                                      
  p &  1 &  e &  4 &  -654.3137    &  -651.8816  &   -654.4730  &    -652.0275  \\                                                      
  p &  1 &  e &  5 &  -563.9041    &  -562.5239  &   -563.8107  &    -562.4224  \\                                                      
  p &  1 &  e &  6 &  -532.1888    &  -531.5656  &   -531.6393  &    -531.0251  \\                                                      
  p &  1 &  e &  7 &  -295.9271    &  -295.1356  &   -295.8749  &    -295.0942  \\                                                    
  p &  1 &  e &  8 &  -213.8730    &  -212.6546  &   -213.8144  &    -212.5940  \\                                                                          
  p &  1 &  e &  9 &  -210.1857    &  -209.9754  &   -209.8876  &    -209.6923  \\                                                  
  p &  1 &  e & 10 &  -156.4334    &  -154.5125  &   -156.5453  &    -154.6271  \\                                                      
  p &  1 &  e & 11 &  -105.5803    &  -105.5427  &   -105.5159  &    -105.4853  \\                                                    
  p &  1 &  e & 12 &   -57.0345    &   -57.0589  &    -57.0950  &     -57.1386  \\                                       
  p &  1 &  e & 13 &   -14.0353    &   -14.3679  &    -13.9650  &     -14.3146  \\             
  p &  1 &  e & 14 &    -0.4303    &    -0.4477  &     -0.9542  &      -0.9691  \\     
  p &  1 &  f &  1 & -1155.8701    & -1153.4283  &  -1155.8380  &   -1153.5668  \\                                                                       
  p &  1 &  f &  2 &  -562.8482    &  -561.2379  &   -562.5708  &    -561.1507  \\                                                                       
  p &  1 &  f &  3 &  -213.9034    &  -212.4739  &   -213.6404  &    -212.4074  \\
  p &  1 &  f &  4 &  -156.8473    &  -154.6979  &   -156.7529  &    -154.8036  \\
  \hline
  p &  2 &  e &  1 & -1770.6656    & -1769.2750  &  -1770.8484  &   -1769.4250  \\
  p &  2 &  e &  2 & -1140.2456    & -1138.0124  &  -1140.4174  &   -1138.1568  \\
  p &  2 &  e &  3 & -1041.6720    & -1040.5995  &  -1041.3059  &   -1040.2204  \\
  p &  2 &  e &  4 &  -638.7466    &  -636.3328  &   -638.9022  &    -636.4746  \\
  p &  2 &  e &  5 &  -556.0217    &  -554.3278  &   -555.9838  &    -554.2686  \\
  p &  2 &  e &  6 &  -542.2887    &  -539.7977  &   -542.3536  &    -539.8546  \\
  p &  2 &  e &  7 &  -517.6946    &  -517.0010  &   -517.2002  &    -516.5167  \\
  p &  2 &  e &  8 &  -284.8796    &  -284.0997  &   -284.8154  &    -284.0460  \\
  p &  2 &  e &  9 &  -202.5942    &  -201.5499  &   -202.5247  &    -201.4285  \\
  p &  2 &  e & 10 &  -201.3838    &  -201.0364  &   -201.1450  &    -200.8627  \\
  p &  2 &  e & 11 &  -178.4708    &  -177.3169  &   -178.4829  &    -177.3364  \\
  p &  2 &  e & 12 &  -141.5481    &  -139.6510  &   -141.6852  &    -139.7908  \\
  p &  2 &  e & 13 &   -97.0685    &   -97.0736  &    -96.9718  &     -96.9846  \\
  p &  2 &  e & 14 &   -49.6789    &   -49.7004  &    -49.7621  &     -49.8017  \\
  p &  2 &  e & 15 &   -10.7659    &   -11.0578  &    -10.7122  &     -11.0214  \\
  p &  2 &  e & 16 &     0.8949    &     0.8811  &     -0.0607  &      -0.0614  \\
  p &  2 &  f &  1 & -1139.2873    & -1137.0473  &  -1139.4552  &   -1137.1878  \\
  p &  2 &  f &  2 &  -553.7206    &  -551.7982  &   -553.7281  &    -551.7824  \\
  p &  2 &  f &  3 &  -540.8119    &  -538.4484  &   -540.8711  &    -538.5027  \\
  p &  2 &  f &  4 &  -202.0169    &  -200.8526  &   -201.9473  &    -200.7820  \\
  p &  2 &  f &  5 &  -178.5053    &  -177.3466  &   -178.5158  &    -177.3643  \\
  p &  2 &  f &  6 &  -141.9751    &  -139.9961  &   -142.0884  &    -140.1105  \\
  \hline
  p &  3 &  e &  1 & -1746.1998    & -1744.8130  &  -1746.3885  &   -1744.9677  \\
  p &  3 &  e &  2 & -1116.6749    & -1114.4549  &  -1116.8539  &   -1114.6063  \\
  p &  3 &  e &  3 & -1019.5787    & -1018.5112  &  -1019.2346  &   -1018.1502  \\
  p &  3 &  e &  4 &  -615.4751    &  -613.0921  &   -615.6299  &    -613.2306  \\
  p &  3 &  e &  5 &  -540.0044    &  -538.4192  &   -539.9576  &    -538.3553  \\
  p &  3 &  e &  6 &  -517.7668    &  -515.3157  &   -517.8192  &    -515.3527  \\
  p &  3 &  e &  7 &  -495.8879    &  -495.0750  &   -495.4796  &    -494.6758  \\
  p &  3 &  e &  8 &  -268.3268    &  -267.5780  &   -268.2512  &    -267.5089  \\
  p &  3 &  e &  9 &  -187.9933    &  -187.6534  &   -187.7801  &    -187.4556  \\
  p &  3 &  e & 10 &  -185.8049    &  -184.8122  &   -185.7869  &    -184.7907  \\
  p &  3 &  e & 11 &  -161.5033    &  -160.3553  &   -161.5289  &    -160.3885  \\
  p &  3 &  e & 12 &  -119.8175    &  -117.9565  &   -119.9861  &    -118.1274  \\
  p &  3 &  e & 13 &   -84.7585    &   -84.8310  &    -84.6283  &     -84.7059  \\
  p &  3 &  e & 14 &   -38.6301    &   -38.6611  &    -38.7404  &     -38.7866  \\
  p &  3 &  e & 15 &    -6.1093    &    -6.3201  &     -6.0817  &      -6.3105  \\
  p &  3 &  f &  1 & -1114.7404    & -1112.5063  &  -1114.9114  &   -1112.6495  \\
  p &  3 &  f &  2 &  -535.4256    &  -533.5448  &   -535.4327  &    -533.5329  \\
  p &  3 &  f &  3 &  -514.7125    &  -512.3361  &   -514.7835  &    -512.3976  \\
  p &  3 &  f &  4 &  -184.6875    &  -183.6222  &   -184.6116  &    -183.5464  \\
  p &  3 &  f &  5 &  -161.5924    &  -160.4312  &   -161.6092  &    -160.4550  \\
  p &  3 &  f &  6 &  -120.1374    &  -118.1289  &   -120.2620  &    -118.2533  \\
  \hline
  p &  4 &  e &  1 & -1713.6219    & -1712.2389  &  -1713.8172  &   -1712.4000  \\
  p &  4 &  e &  2 & -1085.3338    & -1083.1320  &  -1085.5230  &   -1083.2935  \\
  p &  4 &  e &  3 &  -990.1572    &  -989.0906  &   -989.8379  &    -988.7540  \\
  p &  4 &  e &  4 &  -584.6058    &  -582.2643  &   -584.7623  &    -582.4036  \\
  p &  4 &  e &  5 &  -517.4308    &  -515.9667  &   -517.3913  &    -515.9132  \\
  p &  4 &  e &  6 &  -486.6758    &  -484.3335  &   -486.7025  &    -484.3382  \\
  p &  4 &  e &  7 &  -466.6264    &  -465.6147  &   -466.3291  &    -465.3303  \\
  p &  4 &  e &  8 &  -246.4042    &  -245.7078  &   -246.3135  &    -245.6232  \\
  p &  4 &  e &  9 &  -169.3926    &  -168.9900  &   -169.2803  &    -168.8895  \\
  p &  4 &  e & 10 &  -164.1346    &  -163.2778  &   -164.1268  &    -163.2709  \\
  p &  4 &  e & 11 &  -138.6821    &  -137.5338  &   -138.7376  &    -137.5967  \\
  p &  4 &  e & 12 &   -91.7646    &   -89.9697  &    -91.9627  &     -90.1686  \\
  p &  4 &  e & 13 &   -69.1954    &   -69.3391  &    -69.0381  &     -69.1905  \\
  p &  4 &  e & 14 &   -24.1706    &   -24.2514  &    -24.2927  &     -24.3849  \\
  p &  4 &  e & 15 &    -0.5398    &    -0.6185  &     -0.5873  &      -0.6800  \\
  p &  4 &  f &  1 & -1082.0685    & -1079.8421  &  -1082.2434  &   -1079.9890  \\
  p &  4 &  f &  2 &  -509.7563    &  -507.9354  &   -509.7653  &    -507.9274  \\
  p &  4 &  f &  3 &  -481.3761    &  -478.9785  &   -481.4615  &    -479.0518  \\
  p &  4 &  f &  4 &  -161.9885    &  -161.0487  &   -161.9064  &    -160.9677  \\
  p &  4 &  f &  5 &  -138.7889    &  -137.6201  &   -138.8190  &    -137.6568  \\
  p &  4 &  f &  6 &   -91.3108    &   -89.2809  &    -91.4461  &     -89.4149  \\
  \hline
  p &  5 &  e &  1 & -1672.9672    & -1671.5887  &  -1673.1707  &   -1671.7579  \\
  p &  5 &  e &  2 & -1046.2995    & -1044.1221  &  -1046.5037  &   -1044.2985  \\
  p &  5 &  e &  3 &  -953.4279    &  -952.3610  &   -953.1398  &    -952.0554  \\
  p &  5 &  e &  4 &  -546.2840    &  -544.0005  &   -546.4520  &    -544.1499  \\
  p &  5 &  e &  5 &  -488.5539    &  -487.2051  &   -488.5522  &    -487.1929  \\
  p &  5 &  e &  6 &  -449.1836    &  -447.0443  &   -449.1750  &    -447.0100  \\
  p &  5 &  e &  7 &  -429.6461    &  -428.3497  &   -429.4835  &    -428.2014  \\
  p &  5 &  e &  8 &  -219.4138    &  -218.8010  &   -219.3111  &    -218.7041  \\
  p &  5 &  e &  9 &  -145.3259    &  -144.8495  &   -145.3482  &    -144.8810  \\
  p &  5 &  e & 10 &  -137.8843    &  -137.1764  &   -137.9049  &    -137.2015  \\
  p &  5 &  e & 11 &  -109.9130    &  -108.7645  &   -110.0192  &    -108.8773  \\
  p &  5 &  e & 12 &   -58.3222    &   -56.8480  &    -58.4990  &     -56.9939  \\
  p &  5 &  e & 13 &   -50.7280    &   -50.7354  &    -50.6148  &     -50.6655  \\
  p &  5 &  e & 14 &    -7.7992    &    -8.0052  &     -7.8771  &      -8.0924  \\
  p &  5 &  f &  1 & -1041.3209    & -1039.1042  &  -1041.5006  &   -1039.2556  \\
  p &  5 &  f &  2 &  -476.9944    &  -475.2406  &   -477.0092  &    -475.2396  \\
  p &  5 &  f &  3 &  -440.6849    &  -438.2696  &   -440.7871  &    -438.3578  \\
  p &  5 &  f &  4 &  -134.2851    &  -133.4836  &   -134.1997  &    -133.4006  \\
  p &  5 &  f &  5 &  -109.9574    &  -108.7858  &   -110.0083  &    -108.8427  \\
  p &  5 &  f &  6 &   -55.7294    &   -53.6967  &    -55.8738  &     -53.8386  \\
  \hline
  p &  6 &  e &  1 & -1624.2811    & -1622.9082  &  -1624.4946  &   -1623.0871  \\
  p &  6 &  e &  2 &  -999.6731    &  -997.5272  &   -999.9008  &    -997.7271  \\
  p &  6 &  e &  3 &  -909.4166    &  -908.3473  &   -909.1668  &    -908.0795  \\
  p &  6 &  e &  4 &  -500.7168    &  -498.5152  &   -500.9133  &    -498.6910  \\
  p &  6 &  e &  5 &  -453.5133    &  -452.2657  &   -453.5990  &    -452.3453  \\
  p &  6 &  e &  6 &  -405.5917    &  -403.7023  &   -405.5564  &    -403.6445  \\
  p &  6 &  e &  7 &  -384.6923    &  -383.0749  &   -384.6701  &    -383.0616  \\
  p &  6 &  e &  8 &  -187.8794    &  -187.3813  &   -187.7756  &    -187.2839  \\
  p &  6 &  e &  9 &  -115.4064    &  -114.8647  &   -115.5910  &    -115.0538  \\
  p &  6 &  e & 10 &  -107.4654    &  -106.8872  &   -107.5565  &    -106.9875  \\
  p &  6 &  e & 11 &   -75.1659    &   -74.0324  &    -75.3417  &     -74.2140  \\
  p &  6 &  e & 12 &   -31.8870    &   -32.2321  &    -31.7203  &     -32.0773  \\
  p &  6 &  e & 13 &   -18.6187    &   -17.0825  &    -18.8939  &     -17.3598  \\
  p &  6 &  f &  1 &  -992.5603    &  -990.3554  &   -992.7456  &    -990.5119  \\
  p &  6 &  f &  2 &  -437.3322    &  -435.6497  &   -437.3585  &    -435.6610  \\
  p &  6 &  f &  3 &  -392.6572    &  -390.2308  &   -392.7794  &    -390.3377  \\
  p &  6 &  f &  4 &  -101.9670    &  -101.2987  &   -101.8858  &    -101.2209  \\
  p &  6 &  f &  5 &   -75.0565    &   -73.9038  &    -75.1349  &     -73.9873  \\
  p &  6 &  f &  6 &   -13.7462    &   -11.7420  &    -13.8971  &     -11.8896  \\
  \hline
  p &  7 &  e &  1 & -1567.6184    & -1566.2521  &  -1567.8438  &   -1566.4424  \\
  p &  7 &  e &  2 &  -945.5812    &  -943.4744  &   -945.8489  &    -943.7157  \\
  p &  7 &  e &  3 &  -858.1476    &  -857.0726  &   -857.9439  &    -856.8501  \\
  p &  7 &  e &  4 &  -448.2081    &  -446.1309  &   -448.4584  &    -446.3553  \\
  p &  7 &  e &  5 &  -412.3867    &  -411.2085  &   -412.6467  &    -411.4721  \\
  p &  7 &  e &  6 &  -356.0199    &  -354.3418  &   -355.9894  &    -354.2954  \\
  p &  7 &  e &  7 &  -331.7381    &  -329.8469  &   -331.8450  &    -329.9544  \\
  p &  7 &  e &  8 &  -152.5037    &  -152.1410  &   -152.4271  &    -152.0730  \\
  p &  7 &  e &  9 &   -79.5164    &   -78.9618  &    -79.8244  &     -79.2624  \\
  p &  7 &  e & 10 &   -73.0241    &   -72.5164  &    -73.3096  &     -72.8228  \\
  p &  7 &  e & 11 &   -34.5495    &   -33.4575  &    -34.8290  &     -33.7423  \\
  p &  7 &  e & 12 &   -10.6654    &   -11.1332  &    -10.5609  &     -11.0431  \\
  p &  7 &  f &  1 &  -935.8628    &  -933.6706  &   -936.0548  &    -933.8345  \\
  p &  7 &  f &  2 &  -390.9532    &  -389.3438  &   -391.0000  &    -389.3782  \\
  p &  7 &  f &  3 &  -337.3684    &  -334.9385  &   -337.5178  &    -335.0722  \\
  p &  7 &  f &  4 &   -65.3695    &   -64.8141  &    -65.3031  &     -64.7545  \\
  p &  7 &  f &  5 &   -34.2391    &   -33.1393  &    -34.3546  &     -33.2596  \\
  \hline
  p &  8 &  e &  1 & -1503.0436    & -1501.6846  &  -1503.2832  &   -1501.8888  \\
  p &  8 &  e &  2 &  -884.1788    &  -882.1242  &   -884.5188  &    -882.4371  \\
  p &  8 &  e &  3 &  -799.6406    &  -798.5538  &   -799.4915  &    -798.3855  \\
  p &  8 &  e &  4 &  -389.3092    &  -387.4551  &   -389.6161  &    -387.7220  \\
  p &  8 &  e &  5 &  -365.0837    &  -363.8971  &   -365.6923  &    -364.5247  \\
  p &  8 &  e &  6 &  -300.3470    &  -298.8004  &   -300.3543  &    -298.7974  \\
  p &  8 &  e &  7 &  -271.0221    &  -268.9440  &   -271.2514  &    -269.1667  \\
  p &  8 &  e &  8 &  -114.0220    &  -113.8003  &   -114.0257  &    -113.8135  \\
  p &  8 &  e &  9 &   -39.1590    &   -38.7776  &    -39.2819  &     -38.8927  \\
  p &  8 &  e & 10 &   -33.4480    &   -32.8572  &    -34.3120  &     -33.7369  \\
  p &  8 &  f &  1 &  -871.3179    &  -869.1425  &   -871.5191  &    -869.3142  \\
  p &  8 &  f &  2 &  -338.0542    &  -336.5256  &   -338.1362  &    -336.5939  \\
  p &  8 &  f &  3 &  -274.9388    &  -272.5155  &   -275.1274  &    -272.6879  \\
  p &  8 &  f &  4 &   -24.7657    &   -24.3112  &    -24.7300  &     -24.2795  \\
  \hline
  p &  9 &  e &  1 & -1430.6303    & -1429.2795  &  -1430.8881  &   -1429.5015  \\
  p &  9 &  e &  2 &  -815.6569    &  -813.6665  &   -816.1238  &    -814.1074  \\
  p &  9 &  e &  3 &  -733.9028    &  -732.7961  &   -733.8208  &    -732.6947  \\
  p &  9 &  e &  4 &  -325.5931    &  -324.2662  &   -325.7977  &    -324.4125  \\
  p &  9 &  e &  5 &  -310.5955    &  -309.1171  &   -311.9827  &    -310.5460  \\
  p &  9 &  e &  6 &  -238.3578    &  -236.8709  &   -238.4261  &    -236.9316  \\
  p &  9 &  e &  7 &  -202.8896    &  -200.7004  &   -203.2571  &    -201.0579  \\
  p &  9 &  e &  8 &   -73.0521    &   -72.9597  &    -73.2084  &     -73.1275  \\
  p &  9 &  f &  1 &  -799.0327    &  -796.8756  &   -799.2448  &    -797.0577  \\
  p &  9 &  f &  2 &  -278.8611    &  -277.4165  &   -278.9973  &    -277.5388  \\
  p &  9 &  f &  3 &  -205.5318    &  -203.1258  &   -205.7799  &    -203.3573  \\
  \hline
  p & 10 &  e &  1 & -1350.4602    & -1349.1183  &  -1350.7442  &   -1349.3662  \\
  p & 10 &  e &  2 &  -740.2244    &  -738.3134  &   -740.9264  &    -738.9917  \\
  p & 10 &  e &  3 &  -660.9047    &  -659.7676  &   -660.9306  &    -659.7739  \\
  p & 10 &  e &  4 &  -261.4765    &  -260.8329  &   -261.7910  &    -261.1584  \\
  p & 10 &  e &  5 &  -245.0862    &  -243.1896  &   -247.5509  &    -245.6305  \\
  p & 10 &  e &  6 &  -169.8163    &  -168.3318  &   -169.9986  &    -168.5119  \\
  p & 10 &  e &  7 &  -127.6305    &  -125.3873  &   -128.2454  &    -125.9909  \\
  p & 10 &  e &  8 &   -30.0705    &   -30.0973  &    -30.5068  &     -30.5446  \\
  p & 10 &  f &  1 &  -719.1266    &  -716.9898  &   -719.3554  &    -717.1884  \\
  p & 10 &  f &  2 &  -213.6282    &  -212.2721  &   -213.8525  &    -212.4833  \\
  p & 10 &  f &  3 &  -129.3500    &  -126.9722  &   -129.7011  &    -127.3085  \\
  \hline
  p & 11 &  e &  1 & -1262.6702    & -1261.3384  &  -1262.9486  &   -1261.5800  \\
  p & 11 &  e &  2 &  -658.8099    &  -657.0035  &   -659.2418  &    -657.4080  \\
  p & 11 &  e &  3 &  -580.7735    &  -579.5947  &   -580.8077  &    -579.6082  \\
  p & 11 &  e &  4 &  -198.0492    &  -197.6262  &   -198.1619  &    -197.7348  \\
  p & 11 &  e &  5 &  -171.4865    &  -169.6774  &   -172.8307  &    -171.0057  \\
  p & 11 &  e &  6 &   -94.7340    &   -93.2254  &    -94.9419  &     -93.4286  \\
  p & 11 &  e &  7 &   -46.1455    &   -43.8979  &    -46.6227  &     -44.3591  \\
  p & 11 &  f &  1 &  -631.7675    &  -629.6541  &   -631.9932  &    -629.8488  \\
  p & 11 &  f &  2 &  -142.8919    &  -141.6323  &   -143.0247  &    -141.7520  \\
  p & 11 &  f &  3 &   -46.9259    &   -44.5974  &    -47.1915  &     -44.8447  \\
  \hline
  p & 12 &  e &  1 & -1167.3070    & -1165.9859  &  -1167.6105  &   -1166.2522  \\
  p & 12 &  e &  2 &  -570.8072    &  -569.1205  &   -571.4352  &    -569.7222  \\
  p & 12 &  e &  3 &  -493.3064    &  -492.0719  &   -493.4369  &    -492.1821  \\
  p & 12 &  e &  4 &  -130.8967    &  -130.5866  &   -131.3208  &    -131.0026  \\
  p & 12 &  e &  5 &   -90.8206    &   -89.2291  &    -92.5842  &     -90.9799  \\
  p & 12 &  e &  6 &   -12.9718    &   -11.4241  &    -13.3094  &     -11.7600  \\
  p & 12 &  f &  1 &  -537.0794    &  -534.9920  &   -537.3217  &    -535.2029  \\
  p & 12 &  f &  2 &   -66.7091    &   -65.5547  &    -66.9125  &     -65.7453  \\
  \hline
  p & 13 &  e &  1 & -1064.5107    & -1063.2009  &  -1064.8522  &   -1063.5051  \\
  p & 13 &  e &  2 &  -476.9356    &  -475.3866  &   -477.9066  &    -476.3328  \\
  p & 13 &  e &  3 &  -398.5522    &  -397.2506  &   -398.8232  &    -397.5033  \\
  p & 13 &  e &  4 &   -60.0428    &   -59.8436  &    -60.8527  &     -60.6428  \\
  p & 13 &  e &  5 &    -6.1851    &    -4.8368  &     -8.5661  &      -7.2042  \\
  p & 13 &  f &  1 &  -435.2569    &  -433.1982  &   -435.5294  &    -433.4396  \\
\hline
p & 14 &  e &  1 & -954.4838 & -953.1870 & -954.8103  & -953.4754 \\
p & 14 &  e &  2 & -378.3631 & -376.9692 & -379.0683  & -377.6471 \\
p & 14 &  e &  3 & -296.7130 & -295.3439 & -297.0282  & -295.6395 \\
p & 14 &  f &  1 & -326.5725 & -324.5477 & -326.8346  & -324.7777 \\
\hline
p & 15 &  e &  1 & -837.2689 & -835.9837 & -837.6374  & -836.3158 \\
p & 15 &  e &  2 & -274.3015 & -273.0648 & -275.3233  & -274.0618 \\
p & 15 &  e &  3 & -187.6862 & -186.2493 & -188.2086  & -186.7562 \\
p & 15 &  f &  1 & -211.1955 & -209.2084 & -211.4927  & -209.4736 \\
\hline
p & 16 &  e &  1 & -713.0577 & -711.7883 & -713.5047 & -712.1975 \\
p & 16 &  e &  2 & -165.5621 & -164.4891 & -167.0648 & -165.9642 \\
p & 16 &  e &  3 &  -71.7484 &  -70.2558 &  -72.6480 &  -71.1456 \\
p & 16 &  f &  1 &  -89.4432 &  -87.4997 &  -89.8072 &  -87.8322 \\
\hline
p & 17 &  e &  1 & -582.0194 & -580.7651 & -582.6044 & -581.3130\\
p & 17 &  e &  2 &  -52.5453 &  -51.6334 &  -54.7052 &  -53.7642\\
\hline
p & 18 &  e &  1 & -444.3289 & -443.0900 & -445.1540 & -443.8803 \\
\hline
p & 19 &  e &  1 & -300.0312 & -298.8077 & -301.4026 & -300.1488 \\
\hline
p & 20 &  e &  1 & -149.5301 & -148.3249 & -151.6385 & -150.4075 
\end{longtable}

\end{small}
